\documentclass[10pt, final, journal, letterpaper]{IEEEtran} \IEEEoverridecommandlockouts
\usepackage{cite}

\usepackage[pdftex]{graphicx}
\graphicspath{{./images/}}
\DeclareGraphicsExtensions{.pdf, .png}

\usepackage[cmex10]{amsmath}
\usepackage{amssymb}
\usepackage{algorithm}
\usepackage{algorithmic}
\usepackage{array}
\usepackage{url}

\DeclareMathOperator*{\argmin}{arg\,min}

\usepackage{caption}
\usepackage{subcaption}

\usepackage{balance}
\usepackage{dblfloatfix}
\usepackage{xcolor}

\begin{document}

\title{The iterative reweighted Mixed-Norm Estimate\\ for spatio-temporal MEG/EEG source reconstruction}

\author{Daniel~Strohmeier,~\IEEEmembership{}
        Yousra Bekhti,~\IEEEmembership{}
        Jens~Haueisen,~\IEEEmembership{}
        and~Alexandre~Gramfort~\IEEEmembership{}%

\thanks{Copyright (c) 2016 IEEE. Personal use of this material is permitted. However, permission to use this material for any other purposes must be obtained from the IEEE by sending a request to pubs-permissions@ieee.org.}

\thanks{This work was supported by the German Research Foundation \mbox{(Ha 2899/21-1)}, the European Union (FP7-PEOPLE-2013-IAPP 610950), the EDF and Jacques Hadamard Mathematical Foundation (Gaspard Monge Program for Optimization and operations research), the French National Research Agency (ANR-14-NEUC-0002-01), and the National Institutes of Health (R01 MH106174).}

\thanks{D. Strohmeier is with the Institute of Biomedical Engineering and Informatics, Technische Universi\"at Ilmenau, Ilmenau, Germany; e-mail: daniel.strohmeier@tu-ilmenau.de}

\thanks{Y. Bekhti is with the LTCI, CNRS, T\'el\'ecom ParisTech, Universit\'e Paris-Saclay, Paris, France; e-mail: yousra.bekhti@telecom-paristech.fr}

\thanks{J. Haueisen is with the Institute of Biomedical Engineering and Informatics, Technische Universi\"at Ilmenau, Ilmenau, Germany and the Biomagnetic Center, Department of Neurology, Jena University Hospital, Jena, Germany; e-mail: jens.haueisen@tu-ilmenau.de}

\thanks{A. Gramfort is with the LTCI, CNRS, T\'el\'ecom ParisTech, Universit\'e Paris-Saclay, Paris, France and the NeuroSpin, CEA Saclay, Bat. 145, Gif-sur-Yvette Cedex, France; e-mail: alexandre.gramfort@telecom-paristech.fr}}

\markboth{}{}

\maketitle

\begin{abstract}
Source imaging based on magnetoencephalography (MEG) and electroencephalography (EEG) allows for the non-invasive analysis of brain activity with high temporal and good spatial resolution. As the bioelectromagnetic inverse problem is ill-posed, constraints are required. For the analysis of evoked brain activity, spatial sparsity of the neuronal activation is a common assumption. It is often taken into account using convex constraints based on the \mbox{$\mathbf{\textit{l}_{1}}$-norm}. The resulting source estimates are however biased in amplitude and often suboptimal in terms of source selection due to high correlations in the forward model. In this work, we demonstrate that an inverse solver based on a block-separable penalty with a Frobenius norm per block and a \mbox{$\mathbf{\textit{l}_{0.5}}$-quasinorm} over blocks addresses both of these issues. For solving the resulting non-convex optimization problem, we propose the iterative reweighted Mixed Norm Estimate (irMxNE), an optimization scheme based on iterative reweighted convex surrogate optimization problems, which are solved efficiently using a block coordinate descent scheme and an active set strategy. We compare the proposed sparse imaging method to the dSPM and the RAP-MUSIC approach based on two MEG data sets. We provide empirical evidence based on simulations and analysis of MEG data that the proposed method improves on the standard Mixed Norm Estimate (MxNE) in terms of amplitude bias, support recovery, and stability.
\end{abstract}


\section{Introduction}
Source imaging with magnetoencephalography (MEG) and electroencephalography (EEG) delivers insights into the active brain with high temporal and good spatial resolution in a non-invasive way~\cite{baillet-etal:2001}. It is based on solving the bioelectromagnetic inverse problem, which is a high dimensional ill-posed regression problem. In order to render its solution unique, constraints have to be imposed reflecting a priori assumptions on the neuronal sources. 
In the past, several source reconstruction techniques have been proposed, which are based on the assumption that only a few focal brain regions are involved in a specific cognitive task. Inverse methods favoring sparse focal source configurations to explain the MEG/EEG signals include parametric \cite{scherg-etal:85}, scanning \cite{mosher-etal:1992,mosher-leahy:1998,mosher-leahy:1999,xu-etal:2004}, and imaging approaches \cite{Matsuura-Okabe:1995, Ou-etal:2009, gramfort-etal:2012, Lucka-etal:2012, Wipf-Nagarajan:2009, Sorrentino-etal:2009}. These techniques, which are partly used in clinical routine, are suitable e.g. for analyzing evoked responses or epileptic spike activity. Classic MEG/EEG source imaging technique using sparsity-inducing penalties are the Selective Minimum Norm Method \cite{Matsuura-Okabe:1995} or Minimum Current Estimate (MCE) \cite{Uutela-etal:1999}. Both approaches are based on the Lasso~\cite{Tibshirani:1994}, i.e., regularized regression with an \mbox{$\textit{l}_{1}$-norm} penalty, which is a convex surrogate for the optimal, but NP hard \mbox{$\textit{l}_{0}$-norm} regularized regression problem. To reduce the sensitivity to noise and avoid discontinuous, scattered source activations \cite{Ou-etal:2009}, mixed norms such as the \mbox{$\textit{l}_{2,1}$-mixed-norm} used in Group Lasso~\cite{Yuan06GroupLasso} or Group Basis Pursuit\cite{liao-etal:2014} can be applied. The idea is to take the spatio-temporal characteristics of neuronal activity into account by imposing structured sparsity in space or time \cite{haufe-etal:2008,bolstad-etal:2009,Ou-etal:2009,gramfort-etal:2012}. We refer to \cite{huang-etal:2012} for a general review on group selection in high-dimensional models. A prominent example is the Mixed-Norm Estimate (MxNE) proposed in \cite{gramfort-etal:2012}, which extends the MCE to multiple measurement vector problems by applying a block-separable convex penalty. Each block represents the source activation over time of a dipole with free orientation at a specific source location. Spatial sparsity is promoted by an \mbox{$\textit{l}_{1}$-norm} penalty over blocks, whereas a Frobenius norm per block promotes stationary source estimates, i.e., a source with a non-zero amplitude at one time instant has a non-zero amplitude during the full time window of interest~\cite{gramfort-etal:2012, gramfort-etal:2013}. The Frobenius norm also prevents the orientations of the free orientation dipoles from being biased towards the coordinate axes \cite{chang-etal:2013}. These convex approaches allow for fast algorithms with guaranteed global convergence. However, the resulting source estimates are biased in amplitude and often suboptimal in terms of support recovery~\cite{Candes}, which is impaired by the high spatial correlation of the MEG/EEG forward model. As shown e.g. in the field of compressed sensing, promoting sparsity by applying non-convex penalties, such as logarithmic or \mbox{$\textit{l}_{p}$-quasinorm} penalties with $0 < p < 1$, improves support reconstruction in terms of feature selection, amplitude bias, and stability~\cite{Candes, chartrand:2007a, saab-etal:2008}. Several approaches for solving the resulting non-convex optimization problem have been proposed including generalized shrinkage\cite{woodworth-chartrand:2015}, iterative reweighted $\textit{l}_{1}$ \cite{Candes, Gasso, Rakotomamonjy, zhang-rao:2011}, or iterative reweighted $\textit{l}_{2}$ optimization \cite{gorodnitsky-rao:1997, rao-etal:1999, chartrand-yin:2008, cotter-etal:2005, zhang-rao:2011c}. See \cite{wipf-nagarajan:2010, Rakotomamonjy} for a review of these approaches for single and multiple measurement vector problems. 
Several MEG/EEG sparse source imaging techniques based on iterative reweighted $\textit{l}_{2}$ optimization have been proposed \cite{gorodnitsky-etal:1995,gorodnitsky-rao:1997, portniaguine-zhdanov:1999, nagarajan-etal:2006, liu-etal:2005}. An iterative reweighted $\textit{l}_{1}$ optimization technique for EEG source imaging was proposed in \cite{xu-etal:2007}, which however does not impose structured sparsity and applies a fixed orientation constraint \cite{Lin-etal:2006}.
In this paper, we propose the iterative reweighted Mixed-Norm Estimate (irMxNE), a novel MEG/EEG sparse source imaging approach based on the framework of iterative reweighted $\textit{l}_{1}$, which promotes structured sparsity to improve MEG/EEG source reconstruction. A preliminary version of this method was presented in \cite{Strohmeier-etal:2014}. Similar approaches have recently been proposed in other fields of research \cite{wipf-nagarajan:2010, zhang-rao:2011b}. The irMxNE is based on a non-convex block-separable penalty, which combines a Frobenius norm per block and an \mbox{$\textit{l}_{0.5}$-quasinorm} over blocks. The non-convex objective function is minimized iteratively by computing a sequence of weighted MxNE problems. For solving the convex surrogate problems, we propose a new computationally efficient strategy, which combines block coordinate descent~\cite{tseng, Rakotomamonjy, BCD_MMV} and a forward active set strategy with convergence controlled by means of the duality gap, which converges significantly faster than the original MxNE algorithm proposed in \cite{gramfort-etal:2012}. We provide information on the integration of different source orientation constraints~\cite{Lin-etal:2006} and discuss specific problems of MEG/EEG source imaging such as depth bias compensation and amplitude bias correction. We present empirical evidence using simulations and analysis of two experimental MEG data sets that the proposed method outperforms MCE and MxNE in terms of amplitude bias, active source identification, and stability. Finally, we compare the proposed approach with the dSPM~\cite{Dale-etal:2000} and RAP-MUSIC method~\cite{mosher-leahy:1999} based on two MEG data sets.\\

\emph{\textbf{Notation:}}
We mark vectors with bold letters, $\mathbf{a} \in \mathbb{R}^{N}$, and matrices with capital bold letters, $\mathbf{A} \in \mathbb{R}^{N\times M}$. The transpose of a vector or matrix is denoted by $\mathbf{a}^T$ and $\mathbf{A}^T$. The scalar $\mathbf{a}[i]$ is the i$^{\text{th}}$ element of $\mathbf{a}$. $\mathbf{A}[i, :]$ corresponds to the i$^{\text{th}}$ row and $\mathbf{A}[:, j]$ to the j$^{\text{th}}$ column of $\mathbf{A}$. $\|\mathbf{A}\|_{\text{Fro}}$ indicates the Frobenius norm, and $\|\mathbf{A}\|$ the spectral norm of a matrix.

\section{Materials and Methods}
\subsection{The inverse problem}
The MEG/EEG forward problem describes the linear relationship between the MEG/EEG measurements \mbox{$\mathbf{M}\in\mathbb{R}^{N\times T}$} ($N$ number of sensors, $T$ number of time instants) and the source activation \mbox{$\mathbf{X}\in\mathbb{R}^{(SO)\times T}$} ($S$ number of source locations, $O$ number of orthogonal dipoles per source location with $O=1$ if source orientation is postulated, e.g. using the cortical constraint~\cite{Dale:1993}, and typically $O=3$ otherwise). The model then reads:
\begin{equation}
    \mathbf{M} = \mathbf{G} \mathbf{X} + \mathbf{E} \enspace ,
    \label{eqn:forward_model}
\end{equation}
where $\mathbf{G}\in\mathbb{R}^{N\times (SO)}$ is the gain or leadfield matrix, a known instantaneous mixing matrix, which links source and sensor signals. $\mathbf{E}$ is the measurement noise, which is assumed to be additive, white, and Gaussian, $\mathbf{E}[:, j] \sim \mathcal{N}(0, \mathbf{I})$ for all $j$. This assumption is acceptable on the basis of a proper spatial whitening of the data using an estimate of the noise covariance~\cite{engemann-etal:14}. As $SO \gg N$, the MEG/EEG inverse problem is ill-posed and constraints have to be imposed on the source activation matrix $\mathbf{X}$ to render the solution unique. By partitioning $\mathbf{X}$ into S blocks $\mathbf{X}_{s}\in\mathbb{R}^{O\times T}$, where each $\mathbf{X}_{s}$ represents the source activation at a specific source location $s$ over time and across $O$ orthogonal current dipoles, we can apply a penalty term $\mathcal{P}(\mathbf{X})$ promoting block sparsity by combining a Frobenius norm per block and a \mbox{$\textit{l}_{0.5}$-quasinorm} penalty over blocks. The optimization problem reads:
\begin{equation}
\begin{split}
	\mathbf{\widehat{X}} &= \argmin_{\mathbf{X}\in\mathbb{R}^{SO\times T}} \frac{1}{2} \| \mathbf{M} - \mathbf{G} \mathbf{X} \|^2_{\text{Fro}} + \mathcal{P}(\mathbf{X})\\
	\mathbf{\widehat{X}} &= \argmin_{\mathbf{X}\in\mathbb{R}^{SO\times T}} \frac{1}{2} \| \mathbf{M} - \mathbf{G} \mathbf{X} \|^2_{\text{Fro}} + \lambda \sum_{s=1}^S \mathcal{P}_s(\mathbf{X}_{s})\\
	\mathbf{\widehat{X}} &= \argmin_{\mathbf{X}\in\mathbb{R}^{SO\times T}} \frac{1}{2} \| \mathbf{M} - \mathbf{G} \mathbf{X} \|^2_{\text{Fro}} + \lambda \sum_{s=1}^S \sqrt{\|\mathbf{X}_{s}\|_{\text{Fro}} } \enspace ,
\end{split}
\label{eqn:optim_pb}
\end{equation}
where $\lambda > 0$ is the regularization parameter balancing the data fit and penalty term. Similar to the constraint applied in MxNE \cite{gramfort-etal:2012}, $\mathcal{P}(\mathbf{X})$ promotes source estimates with only a few focal sources that have non-zero activations during the entire time interval of interest. The Frobenius norm per block $\mathbf{X}_s$, which combines \mbox{$\textit{l}_2$-norm} penalties over time and orientation as proposed in \cite{Uutela-etal:1999, Ou-etal:2009, gramfort-etal:2013}, imposes stationarity of the source estimate and prevents the source orientations from being biased towards the coordinate axes \cite{chang-etal:2013}. The \mbox{$\textit{l}_{0.5}$-quasinorm} penalty promotes spatial sparsity.

\subsection{Iterative reweighted Mixed Norm Estimate}
The proposed block-separable regularization functional is an extension of the \mbox{$\textit{l}_{2, p}$-quasinorm} penalty with $0<p<1$ used in \cite{Gasso, Candes, Rakotomamonjy, cotter-etal:2005}. These works showed, based on the framework of Difference of Convex functions programming or Majorization-Minimization algorithms, that the resulting non-convex optimization problem can be solved by iteratively solving a sequence of weighted convex surrogate optimization problems with weights being defined based on the previous estimate. The convex surrogate problem is obtained by replacing the non-decreasing concave function $\mathcal{P}_s(\mathbf{X}_{s})$ with a convex upper bound using a local linear approximation at the current estimate. By solving this sequence of surrogate problems, the value of the non-convex objective function decreases, but without guarantee for convergence to a global minimum.
The cost function in Eq.~\eqref{eqn:optim_pb} can thus be minimized by computing the sequence of convex problems given in Eq.~\eqref{eqn:optim_irmxne}. The weights for the \mbox{k$^{\text{th}}$} iteration are obtained from the previous source estimate $\mathbf{\widehat{X}}^{(k-1)}$.
Intuitively, sources with high amplitudes in the \mbox{(k-1)$^{\text{th}}$} iteration will be less penalized in the \mbox{k$^{\text{th}}$} iteration and therefore further promoted.
\begin{equation}
\begin{split}
    \mathbf{\widehat{X}}^{(k)} &=
            \argmin_{\mathbf{X}\in\mathbb{R}^{SO\times T}} \frac{1}{2} \| \mathbf{M} - \mathbf{G} \mathbf{X} \|^2_{\text{Fro}}
            + \lambda \sum_{s}{ \frac{\|\mathbf{X}_{s}\|_{\text{Fro}}}{2 \sqrt{\left\|\mathbf{\widehat{X}}^{(k-1)}_{s}\right\|_{\text{Fro}}}}}\\
                                &=
            \argmin_{\mathbf{X}\in\mathbb{R}^{SO\times T}} \frac{1}{2} \| \mathbf{M} - \mathbf{G} \mathbf{X} \|^2_{\text{Fro}}
            + \lambda \sum_{s}{{\frac{1}{\mathbf{w}^{(k)}[s]}}\|\mathbf{X}_{s}\|_{\text{Fro}}}
\end{split}
    \label{eqn:optim_irmxne}
\end{equation}
As each iteration is equivalent to solving a weighted MxNE problem, we call this optimization scheme the iterative reweighted MxNE (irMxNE). Due to the non-convexity of the optimization problem in Eq.~\eqref{eqn:optim_pb}, the procedure is sensitive to the initialization of $\mathbf{w}^{(k)}[s]$. In this paper, we use $\mathbf{w}^{(1)}[s]=1$ for all $s$ as proposed in \cite{Gasso}. Consequently, the first iteration of irMxNE is equivalent to solving a standard MxNE problem. As each iteration of the iterative scheme in Eq.~\eqref{eqn:optim_irmxne} solves a convex problem with guaranteed global convergence, the initialization of $\mathbf{X}$ has no influence on the final solution. $\mathbf{X}$ can thus be chosen arbitrarily and we use warm starts for improving the computation time. For sources with $\|\mathbf{\widehat{X}}^{(k)}_{s}\|_{\text{Fro}}=0$, Eq.~\eqref{eqn:optim_irmxne} has an infinite regularization term. Typically, a smoothing parameter $\epsilon$ is added to avoid weights to become zero~\cite{Gasso, Candes, chartrand-yin:2008}. Here, we reformulate the weighted MxNE subproblem and apply the weights without epsilon smoothing by scaling the gain matrix with a weighting matrix $\mathbf{W}^{(k)}$ as given in Eq.~\eqref{eqn:optim_irmxne_final}. After convergence, we reapply the scaling to $\mathbf{\widetilde{X}}^{(k)}$ to obtain the final estimate $\mathbf{\widehat{X}}^{(k)}$.
\begin{equation}
\begin{split}
\mathbf{\widetilde{X}}^{(k)} &= \argmin_{\mathbf{X}\in\mathbb{R}^{SO\times T}} \frac{1}{2} \| \mathbf{M} - \mathbf{G} \mathbf{W}^{(k)} \mathbf{X} \|^2_{\text{Fro}} + \lambda \sum_{s}{{\|\mathbf{X}_{s}\|_{\text{Fro}}}}\\
&= \argmin_{\mathbf{X}\in\mathbb{R}^{SO\times T}} \frac{1}{2} \| \mathbf{M} - \mathbf{G}^{(k)} \mathbf{X} \|^2_{\text{Fro}} + \lambda \sum_{s}{{\|\mathbf{X}_{s}\|_{\text{Fro}}}}\\
\mathbf{\widehat{X}}^{(k)} &= \mathbf{W}^{(k)}\mathbf{\widetilde{X}}^{(k)}
\end{split}
\label{eqn:optim_irmxne_final}
\end{equation}
with $\mathbf{W}^{(k)}\in\mathbb{R}^{SO\times SO}$ being a diagonal matrix, which is computed according to Eq.~\eqref{eqn:comp_weights}:
\begin{equation}
\begin{split}
\mathbf{W}^{(k)} &= \text{diag}({\mathbf{w}^{(k)}} \otimes \mathbf{1}_{(O)})\\
\text{with } \mathbf{w}^{(k)}[s] &= 2 \sqrt{\left\|\mathbf{\widehat{X}}^{(k-1)}_{s}\right\|_{\text{Fro}}},
\end{split}
\label{eqn:comp_weights}
\end{equation}
where $\mathbf{1}_{(O)} \in \mathbb{R}^O$ is a vector of ones and $\otimes$ is the Kronecker product. 
In each MxNE iteration, we restrict the source space to source locations with $\mathbf{w}^{(k)}[s]>0$ to reduce the computation time.\\

We control the global convergence of each weighted MxNE subproblems in Eq.~\eqref{eqn:optim_irmxne_final} by monitoring the duality gap. For details on convex duality in the context of optimization with sparsity-inducing penalties, we refer to \cite{bach-etal:2012}. In the following, we summarize the rationale for this stopping criterion. For a general minimization problem, the minimum of the primal objective function $\mathcal{F}_p(\mathbf{X})$ is bounded below by the maximum of the associated dual objective function $\mathcal{F}_d(\mathbf{Y})$, i.e., $\mathcal{F}_p(\mathbf{X}^*) \geq \mathcal{F}_d(\mathbf{Y}^*)$, where $\mathbf{X}^*$ and $\mathbf{Y}^*$ are the optimal solutions of the primal and dual problem. The duality gap $\eta = \mathcal{F}_p(\mathbf{X}) - \mathcal{F}_d(\mathbf{Y})\geq 0$, where $\mathbf{X}$ and $\mathbf{Y}$ are the current values of the primal and dual variable, is thus non-negative and provides an upper bound on the difference between $\mathcal{F}_p(\mathbf{X})$ and $\mathcal{F}_p(\mathbf{X}^*)$. If strong duality holds, the duality gap at the optimum is zero. To use this stopping criterion in practice, we need to derive the dual problem and choose a good feasible dual variable $\mathbf{Y}$ given a value of $\mathbf{X}$, which allows for $\eta=0$ at the optimum.\\ 

Due to Slater's conditions \cite{Boyd_Vandenberghe04}, strong duality holds for the MxNE subproblem and we can check convergence of an iterative optimization scheme solving Eq.~\eqref{eqn:optim_irmxne_final} by computing the current duality gap $\eta^{(i)} = \mathcal{F}_p(\mathbf{X}^{(i)}) - \mathcal{F}_d(\mathbf{Y}^{(i)}) \geq 0$. Based on the Fenchel-Rockafellar duality theorem \cite{rockafellar:1997}, the dual objective function associated to the primal objective function
\begin{equation*}
\begin{split}
\mathcal{F}_p\left(\mathbf{X}\right) &= \frac{1}{2} \| \mathbf{M} - \mathbf{G} \mathbf{X}\|^2_{\text{Fro}} + \lambda \Omega(\mathbf{X})\\
&= \frac{1}{2} \| \mathbf{M} - \mathbf{G} \mathbf{X}\|^2_{\text{Fro}} + \lambda \sum_{s}{{\|\mathbf{X}_{s}\|_{\text{Fro}}}}
\end{split}
\label{eqn:primal_problem}
\end{equation*} 
is given in Eq.~\eqref{eqn:dual_problem}. For a detailed derivation, we refer to \cite{gramfort-etal:2012}.
\begin{equation}
\mathcal{F}_d\left(\mathbf{Y}\right) = -\frac{1}{2}\| \mathbf{Y}\|_\text{Fro}^2 + \mathrm{Tr}\left(\mathbf{Y}^{T}\mathbf{M}\right) - \lambda
\Omega^*\left({\mathbf{G}^T\mathbf{Y}}/{\lambda}\right)
\label{eqn:dual_problem}
\end{equation}
where $\mathrm{Tr}$ indicates the trace of a square matrix, and $\Omega^*$ the Fenchel conjugate of $\Omega$, which is the indicator function of the associated dual norm. As shown in \cite{gramfort-etal:2012}, a natural mapping from the primal to the dual space is given by a scaling of the residual $\mathbf{\widetilde{Y}} = \mathbf{M}-\mathbf{G}\mathbf{X}$. This is motivated by the fact that the solution of the dual problem at the optimum is proportional to the residual, which follows from the associated KKT conditions \cite{gramfort-etal:2012}. The scaling is done according to Eq.~\eqref{eqn:dual_variable} such that the dual variable $\mathbf{Y}$ satisfies the constraint of $\Omega^*$.
\begin{equation}
\mathbf{Y} = \left.\mathbf{\widetilde{Y}} \middle/ \max\right.\left({\max\limits_s \left\|\mathbf{G}_s^{T}\mathbf{\widetilde{Y}}\right\|_{\text{Fro\ }}} \middle/ \lambda,\ 1\right)
\label{eqn:dual_variable}
\end{equation}

In practice, we terminate the iterative optimization scheme for solving MxNE, when the estimate at the \mbox{i$^{\text{th}}$} inner iteration $\mathbf{{X}}^{(i)}$ is $\epsilon$-optimal with $\epsilon=10^{-6}$, i.e., $\eta^{(i)}<10^{-6}$. According to \cite{gramfort-etal:2012}, this is a conservative choice provided that the data is scaled by spatial pre-whitening.\\

For solving the weighted MxNE subproblems, we propose a block coordinate descent (BCD) scheme \cite{tseng}, which, for the problem at hand, converges faster than the Fast Iterative Shrinkage-Thresholding algorithm (FISTA) proposed earlier in~\cite{gramfort-etal:2012} (cf. section~\ref{sec:realdata}). A BCD scheme for solving the Group LASSO was proposed in \cite{Rakotomamonjy, BCD_MMV}. The subproblem per block has a closed form solution, which involves applying the group soft-thresholding operator, the proximity operator associated to the \mbox{$\textit{l}_{2,1}$-mixed-norm}~\cite{gramfort-etal:2012}. Accordingly, the closed form solution for the BCD subproblems solving the MxNE problem can be derived, which is given in Eq.~\eqref{eqn:subBCDsolution}.
\begin{equation}
\begin{split}
\mathbf{\overline{X}}_{s}^{(k)} &= \mathbf{X}_{s}^{(k-1)} + \boldsymbol{\mu}[s]\mathbf{G}_{s}^T \left(\mathbf{M} - \mathbf{GX}^{(k-1)}\right)\\
\mathbf{\widetilde{X}}_{s}^{(k)} &= \mathbf{\widetilde{X}}_{s}^{(k)}\max\left( 1 - {{\boldsymbol{\mu}[s] \lambda}\over{\max\left(\left\| \mathbf{\overline{X}}_{s}^{(k)} \right\|_{\text{Fro}},\,\boldsymbol{\mu}[s] \lambda\right)}},\,0\right)
\end{split}
\label{eqn:subBCDsolution}
\end{equation}\\

The step length $\boldsymbol{\mu}[s]$ for each BCD subproblem is determined by $\boldsymbol{\mu}[s] = L_s^{-1}$ with $L_s = \| \mathbf{G}_{s}^T \mathbf{G}_{s}\|$ being the Lipschitz constant of the data-fit restricted to the s$^\text{th}$ source location. This step length is typically larger than the step length applicable in iterative proximal gradient methods, which is upper-bounded by the inverse of $L = \|\mathbf{G}^T \mathbf{G}\|$. Pseudo code for the BCD scheme is shown in \mbox{Algorithm~\ref{alg:mxne}}.
\begin{algorithm}
\caption{MxNE with BCD}
\label{alg:mxne}
\begin{algorithmic}[1]
\REQUIRE $\mathbf{M}$, $\mathbf{G}$, $\mathbf{X}$, $\boldsymbol{\mu}$, $\lambda > 0$, $\epsilon > 0$, and $S$.
\STATE Initialization: $\eta = \mathcal{F}_p\left(\mathbf{X}\right) - \mathcal{F}_d\left(\mathbf{Y}\right)$
\WHILE{$\eta \geq \epsilon$}
    \FOR{$s=1$ to $S$}
	    \STATE $\mathbf{X}_{s}$ $\longleftarrow$ Solve Eq.~\eqref{eqn:subBCDsolution} with $\mathbf{X}$, $\boldsymbol{\mu}$, and $\mathbf{M}$
    \ENDFOR
    \STATE $\eta = \mathcal{F}_p(\mathbf{X}) - \mathcal{F}_d(\mathbf{Y})$
\ENDWHILE
\end{algorithmic}
\end{algorithm}

The BCD scheme is typically applied using a cyclic sweep pattern, i.e., all blocks are updated in a cyclic order in each iteration. However, as the penalty term in Eq.~\eqref{eqn:optim_pb} promotes spatial sparsity, most of the blocks of $\mathbf{\widehat{X}}$ are zero. We can thus reduce the computation time by primarily updating blocks, that are likely to be non-zero, while keeping the remaining blocks at zero. For this purpose, data-dependent sweep patterns (such as greedy approaches based on steepest descent \cite{li-osher:2009,wei-etal:2012}) or active set strategies \cite{friedman-etal:2010, roth-etal:08} can be applied. In this paper, we combine BCD with a forward active set strategy proposed in \cite{roth-etal:08, gramfort-etal:2012}. Pseudo code for the proposed MxNE solver is provided in \mbox{Algorithm~\ref{alg:mxne_as}}. 
\begin{algorithm}
\caption{MxNE with BCD and active set strategy}
\label{alg:mxne_as}
\begin{algorithmic}[1]
\REQUIRE $\mathbf{M}$, $\mathbf{G}$, $\lambda > 0$, $\epsilon > 0$, and $S$.
\STATE Initialization: $\mathbf{X}=\mathbf{0}$, $\mathcal{A}=\{ \}$, $\eta = \mathcal{F}_p\left(\mathbf{X}\right) - \mathcal{F}_d\left(\mathbf{Y}\right)$
\FOR{$s=1$ to $S$}
    \STATE $\boldsymbol{\mu}[s] = \| \mathbf{G}_{s}^T \mathbf{G}_{s}\|^{-1}$
\ENDFOR
\WHILE{$\eta \geq \epsilon$}
    \STATE $\mathcal{A}^* \subseteq \{s\ |\ \| \mathbf{G}_{s}^T (\mathbf{M}-\mathbf{G}\mathbf{X}) \|_{\text{Fro}} > \lambda\}$
    \STATE $\mathcal{A} = \mathcal{A}\ \cup\ \mathcal{A}^*$
    \STATE Define $\mathbf{G^{\mathcal{A}}}$ and $\mathbf{X^{\mathcal{A}}}$ by restricting $\mathbf{G}$ and $\mathbf{X}$ to $\mathcal{A}$
    \STATE $\mathbf{\widehat{X}}^{\mathcal{A}}$ $\longleftarrow$ Solve Algorithm~\ref{alg:mxne} with $\boldsymbol{\mu}$, $\mathbf{G}^{\mathcal{A}}$ and $\mathbf{X}_0=\mathbf{X}^{\mathcal{A}}$ 
    \STATE $\mathbf{X} = \mathbf{\widehat{X}}^{\mathcal{A}} \text{\ for\ } s \in \mathcal{A}, \text{else\ } \mathbf{0}$
	\STATE $\eta = \mathcal{F}_p(\mathbf{X}) - \mathcal{F}_d(\mathbf{Y})$
\ENDWHILE
\end{algorithmic}
\end{algorithm}

We start by estimating an initial active set of sources $\mathcal{A}$ by evaluating the Karush-Kuhn-Tucker (KKT) optimality conditions, which state that \mbox{$\widehat{\mathbf{X}}_s=\mathbf{0}$} if \mbox{$\|\mathbf{G}_s^T(\mathbf{M}-\mathbf{G}\mathbf{X})\|_{\text{Fro}}\leq \lambda$ \cite{gramfort-etal:2012}.} We select the N sources as the initial active set, which violate this condition the most (we use $N=10$ in practice). Subsequently, we restrict the source space to the sources in $\mathcal{A}$ and estimate $\mathbf{\widehat{X}}^{\mathcal{A}}$ by solving Eq.~\eqref{eqn:optim_irmxne_final} with convergence controlled by the duality gap. 
After convergence of this restricted optimization problem, we check whether $\mathbf{\widehat{X}}^{\mathcal{A}}$ is also an $\epsilon$-optimal solution for the original optimization problem (without restricting the source space to $\mathcal{A}$) by computing the corresponding duality gap $\eta$ assuming that all sources, which are not part of the active set, have zero activation. If $\mathbf{\widehat{X}}^{\mathcal{A}}$ is not an $\epsilon$-optimal solution indicated by $\eta\geq\epsilon$, we re-evaluate the KKT optimality conditions and update the active set $\mathcal{A}$ by adding the N sources with the highest score. We then repeat the procedure with warm start.\\

We terminate irMxNE when $\|\mathbf{\widehat{X}}^{(k)} - \mathbf{\widehat{X}}^{(k-1)}\|_{\infty} <\tau$ with a user specified threshold $\tau$, which we set to $10^{-6}$ in practice. The proposed optimization algorithm for irMxNE is fast enough to allow its usage in practical MEG/EEG applications. Full pseudo code for irMxNE is provided in \mbox{Algorithm~\ref{alg:irmxne}}.
\begin{algorithm}
\caption{Iterative reweighted MxNE}
\label{alg:irmxne}
\begin{algorithmic}[1]
\REQUIRE $\mathbf{M}$, $\mathbf{G}$, $\lambda > 0$, $\epsilon > 0$, $\tau > 0$, and $K$.
\STATE Initialization: $\mathbf{W}^{(1)}=\mathbf{I}$, $\mathbf{\widehat{X}}^{(1)}$ 
\FOR{$k=1$ to $K$}
    \STATE $\mathbf{G}^{(k)} = \mathbf{G}\mathbf{W}^{(k)}$
    \STATE $\mathbf{\widetilde{X}}^{(k)}$ $\longleftarrow$ Solve Algorithm~\ref{alg:mxne_as} with $\mathbf{G}^{(k)}$ and  $\mathbf{X}^{(k)}$
    \STATE $\mathbf{\widehat{X}}^{(k)} = \mathbf{W}^{(k)}\mathbf{\widetilde{X}}^{(k)}$
    \IF{$\|\mathbf{\widehat{X}}^{(k)} - \mathbf{\widehat{X}}^{(k-1)}\|_{\infty}<\tau$}
            \STATE break
    \ENDIF
    \STATE $\mathbf{W}^{(k+1)}$ $\longleftarrow$ Solve Eq.~\ref{eqn:comp_weights} with $\mathbf{\widehat{X}}^{(k)}$
\ENDFOR
\end{algorithmic}
\end{algorithm}

\subsection{Source constraints and bias}
\subsubsection{Source orientation}
The proposed BCD scheme is applicable for MEG/EEG inverse problems without and with orientation constraint. For imposing a loose orientation constraint~\cite{Lin-etal:2006}, we apply a weighting matrix $\mathbf{K} = \text{diag}([1, \rho, \rho])$ to each block of the gain matrix $\mathbf{G}_s\in\mathbb{R}^{N\times 3}$ with $\mathbf{G}_s[:, 1]$ corresponding to the dipole orientated normally to the cortical surface, and $\mathbf{G}_s[:, 2]$ and $\mathbf{G}_s[:, 3]$ to the two tangential dipoles. The weighting parameter $0 < \rho \leq 1$ controls up to which angle the rotating dipole may deviate from the normal direction~\cite{Lin-etal:2006, gramfort-etal:2013}. The orientation-weighted gain matrix $\mathbf{\widetilde{G}}$ is hence defined as $\mathbf{\widetilde{G}} = \mathbf{G}\left(\mathbf{I}_{(S)}\otimes \mathbf{K}\right)$, where $\mathbf{I}_{(S)} \in\mathbb{R}^{S\times S}$ is the identity matrix. Since the penalty in Eq.~\eqref{eqn:subBCDsolution} does not promote sparsity along orientations, the irMxNE result is not biased towards the coordinate axes~\cite{chang-etal:2013}. When the source orientation is postulated a priori (e.g. normal to the cortical surface), each block $\mathbf{X}_{s}$ corresponds to the activation of a fixed dipole. Consequently, the Frobenius norm per block can be replaced by the \mbox{$\textit{l}_{2}$-norm} of the source activation of the corresponding fixed dipole.
\subsubsection{Depth bias compensation}
Due to the attenuation of the bioelectromagnetic field with increasing distance between source and sensor, deep sources require higher source amplitudes to generate sensor signals of equal strength compared to superficial sources. Consequently, inverse methods, which are based on constraints penalizing the source amplitudes, have a bias towards superficial sources. In order to compensate this bias, each block of the gain matrix is weighted a priori. Here, we apply the depth bias compensation proposed in \cite{huang-etal:2014}, which computes the weights used for depth bias compensation based on the SVD of the gain matrix.
\subsubsection{Amplitude bias compensation}
Source activation estimated with source reconstruction approaches based on \mbox{$\textit{l}_{p}$-quasinorms} with $0 < p \leq 1$, such as MxNE and irMxNE, show a varying degree of amplitude bias due to the inherent shrinkage. The standard practice for compensating the amplitude bias consists in computing the least squares fit after restricting the source space to the support of $\mathbf{\widehat{X}}$, which is typically an over-determined optimization problem. In contrast, we apply the debiasing approach proposed in \cite{gramfort-etal:2013}, which preserves the source characteristics and orientations estimated with irMxNE by estimating a scaling factor for each source, which is constrained to be above 1 and constant over orientation and time. The bias corrected source estimate $\mathbf{\widetilde{X}}$ is computed using $\mathbf{D}$ as $\mathbf{\widetilde{X}} = \mathbf{D} \mathbf{\widehat{X}}$, where the diagonal scaling matrix $\mathbf{D}$ is estimated based on the convex problem:
\begin{equation*}
     \mathbf{\widehat{D}} = \argmin_{\mathbf{D}}  \| \mathbf{M} - \mathbf{G} (\mathbf{D}\otimes \mathbf{I}_{(O)}) \mathbf{\widehat{X}}
\|^2_{\text{Fro}} \enspace \textrm{s.t.} \enspace
    \begin{cases}
        {\mathbf{D_{ij}} \ge 1,} & {i = j}\\
        {\mathbf{D_{ij}} = 0,}& {i \neq j}
    \end{cases}
\label{eqn:debias}
\end{equation*}

\subsection{Simulation setup}
We compare MCE, MxNE and irMxNE in terms of amplitude bias, support recovery, and stability using simulated auditory evoked fields. The simulation, which was repeated 100 times, is based on a real gain matrix computed with a three-shell boundary element model using 4699 cortical sources with fixed orientation (normal to the cortical surface), and a 306-channels Elekta Neuromag Vectorview system (Elekta Neuromag Oy, Helsinki, Finland) with 102 magnetometers and 204 gradiometers. The sampling rate was set to 1\,kHz and we restricted the analysis to the time window from \mbox{60\,ms to 150\,ms}. We generated single trials by activating two dipolar sources, one in each transverse temporal gyrus, with Gaussian functions peaking at 100\,ms and 110\,ms with a peak amplitude of 55\,nAm and 45\,nAm, $\mathbf{X}_{\text{sim}}$. Background activity was generated by ten dipolar sources placed randomly on the cortical surface. Each dipole was activated with filtered white noise with a peak amplitude of 100\,nAm. The filter coefficients were determined by fitting an auto-regressive process of order 5 to real baseline MEG data~\cite{gramfort-etal:2013}. By averaging 100 single trials, the SNR of the evoked response, which we compute using spatial whitened data as \mbox{$\textrm{SNR}=\|\mathbf{M}_{\textrm{signal}}\|^2_{\textrm{Fro}} / \|\mathbf{M}_{\textrm{noise}}\|^2_{\textrm{Fro}}$}, was set to \mbox{$\textrm{SNR}=2.63\pm 0.46$}. Source reconstruction was computed without orientation constraint, where none of the dipoles used to generate the gain matrix was oriented perpendicularly to the cortical surface. All methods were applied with different regularization parameters $\lambda$ given as a percentage of the respective $\lambda_{\text{max}}$, which is the smallest regularization parameter leading to an empty active set~\cite{gramfort-etal:2012}. We evaluate the source reconstruction performance by means of the true and false positives counts. We consider a source to be a true positive, if its geodesic distance along the cortical surface from the true source location is less than 1\,cm. A value of 1\,cm is what would be considered an acceptable localization error for most neuroscience applications. Moreover, we present the active set size and the root mean square error in the sensor space, \mbox{RMSE = $\|\mathbf{G}\mathbf{{X}}_\text{sim} - \mathbf{G}\mathbf{\widehat{X}}\|_{\text{Fro}}$}. To evaluate the stability of the reconstructed support, we compute Krippendorff's alpha~\cite{hayes-krippendorff:2007}.

\subsection{Experimental MEG data}
We evaluate the performance of MxNE and irMxNE using data from the MIND multi-site MEG study~\cite{aine-etal:2012, weisend-etal:2007, Ou-etal:2007}. We use two different data sets from one exemplary subject, auditory evoked fields (AEF) and somatosensory evoked fields (SEF), recorded using the 306-channels Elekta Neuromag Vectorview system. A detailed description of the data and paradigms can be found in \cite{aine-etal:2012, weisend-etal:2007, Ou-etal:2007}.
For the AEF data set, we report results for AEFs evoked by left auditory stimulation with pure tones of 500\,Hz. The analysis window for source estimation was chosen from \mbox{50\,ms to 200\,ms} based on visual inspection of the evoked data to capture the dominant N100m component. For the SEF data set, we analyzed SEFs evoked by bipolar electrical stimulation (0.2\,ms in duration) of the left median nerve. To capture the main peaks of the evoked response and to exclude the strong stimulus artifact, the analysis window was chosen from \mbox{18\,ms to 200\,ms} based on visual inspection. Following the standard pipeline from the MNE software~\cite{mne}, signal preprocessing for both data sets consisted of signal-space projection for suppressing environmental noise, and baseline correction using pre-stimulus data (from -200\,ms to -20\,ms). Epochs with peak-to-peak amplitudes exceeding predefined rejection parameters (3\,pT for magnetometers, 400\,pT/m for gradiometers, and 150\,μV for EOG) were assumed to be affected by artifacts and discarded. This resulted in 96 (AEF) and 294 (SEF) artifact-free epochs, which were resampled to 500\,Hz. The gain matrix was computed using a set of 7498 cortical locations, and a three-layer boundary element model. 
The stability of the source reconstruction was tested using a resampling technique. For each data set, we generated 100 random sets of epochs by randomly selecting 80\% of all available epochs without replacement. The noise covariance matrix for spatial whitening was estimated for each subsample using pre-stimulus data (from -200\,ms to -20\,ms). We applied both MxNE and irMxNE on the average of each random set without orientation constraint. Due to the lack of a ground truth, the source reconstruction performance is evaluated by means of the Goodness of Fit (GOF) and the active set size. To compare results with well established source reconstruction techniques, we compute the dSPM solution~\cite{Dale-etal:2000} without orientation constraint and the RAP-MUSIC estimate~\cite{mosher-leahy:1999} for both data sets using the MNE-Python software~\cite{mne-python}. For RAP-MUSIC, we use single-dipole and two-dipole independent topographies to address the problem of correlated sources \cite{mosher-leahy:1998}. A similar idea is pursued e.g. by dual core beamformers~\cite{diwakar-etal:2011}. The correlation threshold was set to 0.95 as proposed by Mosher et al. \cite{mosher-leahy:1998}. The rank of the signal subspace was determined by thresholding the eigenvalues of the data covariance based on an estimate of the noise variance. As we apply a spatial whitening, the threshold was set to 1.

\section{Results}
\subsection{Simulation study}
\begin{figure*}[!t]
    \centering
    \includegraphics[width=0.94\linewidth]{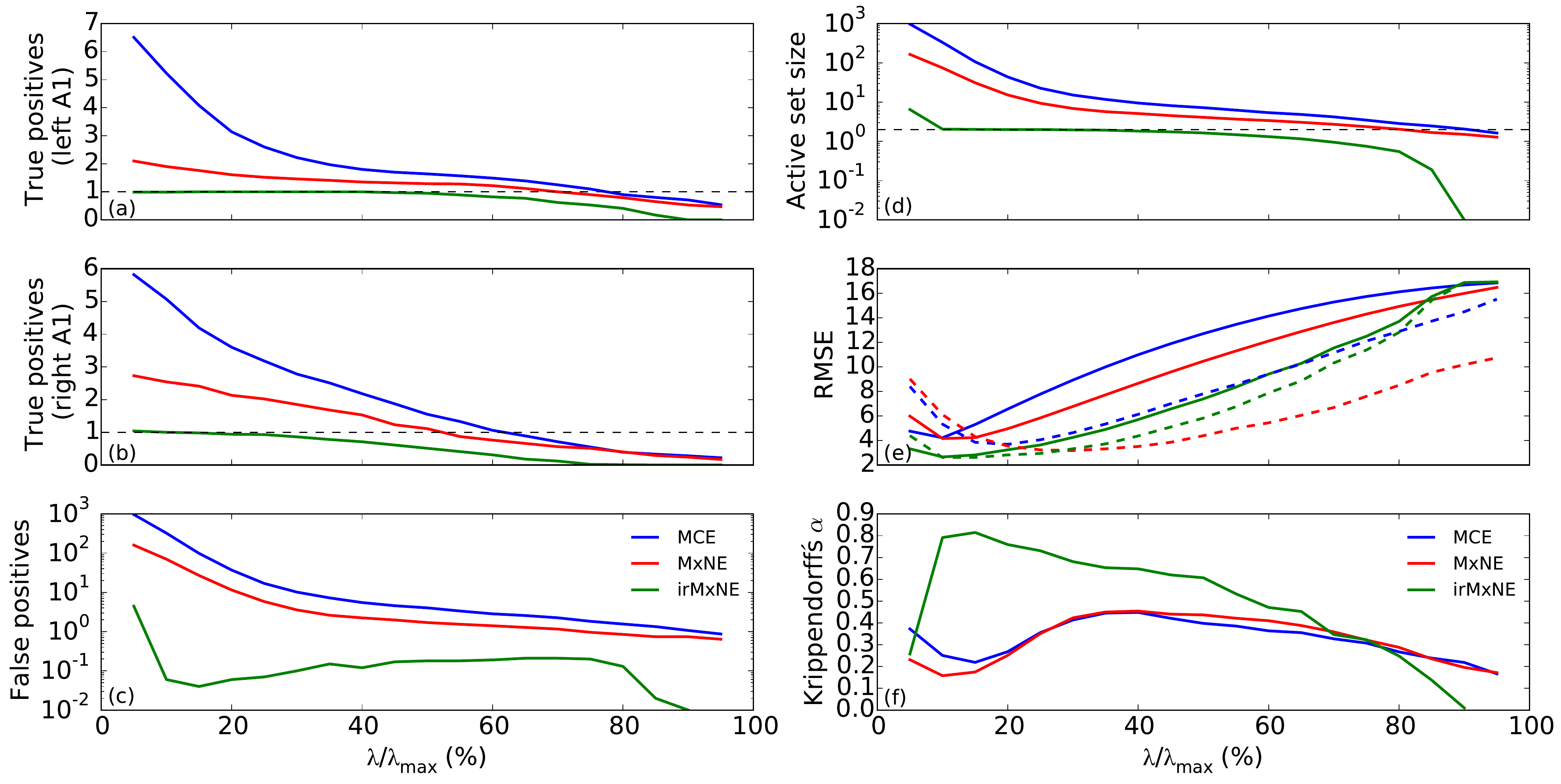}
    \caption{Results of the simulation study based on simulated AEFs for MCE, MxNE and irMxNE. The source space contained a total of 4699 sources and the simulation was repeated 100 times:
    (a) mean true positive count (left A1), (b) mean true positive count (right A1), (c) mean false positive count, (d) mean active set size, (e) mean RMSE without (solid) and with (dashed) debiasing, and (f) Krippendorff's $\alpha$.}
    \label{fig:results_simulation}
\end{figure*}

The results of the simulation study (100 repetitions) for different regularization parameters (from 5 to 100\,\% of $\lambda_{\text{max}}$) are presented in Fig.~\ref{fig:results_simulation}.
The source space contained 4699 sources and one source per hemisphere was active indicated by the horizontal dashed lines in Fig.~\ref{fig:results_simulation}a and b. True positive counts above this threshold indicate suboptimally sparse source estimates, whereas counts close to zero indicate false negatives. We can see that the irMxNE approach provides the best support recovery. It allows to reconstructs single dipoles in both regions of interest, whereas MCE and MxNE find multiple correlated sources. Particularly for low values of $\lambda$, MCE and MxNE overestimate the size of the active set leading to a large number of false positives, whereas irMxNE generates significantly less false positives. The mean active set size confirms that irMxNE provides the sparsest result of all three methods. The mean RMSE is shown in Fig.~\ref{fig:results_simulation}e.  While all methods profit from the debiasing procedure, the effect on irMxNE is less pronounced compared to the other methods indicating a reduced amplitude bias. The best result is obtained with irMxNE. \mbox{Krippendorff's $\alpha$} indicates that the support reconstructed with irMxNE is more stable compared to MCE or MxNE. The source estimate is thus less dependent on the epochs used for generating the evoked response.

\subsection{Experimental MEG data}
\label{sec:realdata}
\subsubsection{Auditory evoked fields}
\label{sec:realdata_aud}
\begin{figure}[!h]
\centering
\includegraphics[width=0.94\columnwidth]{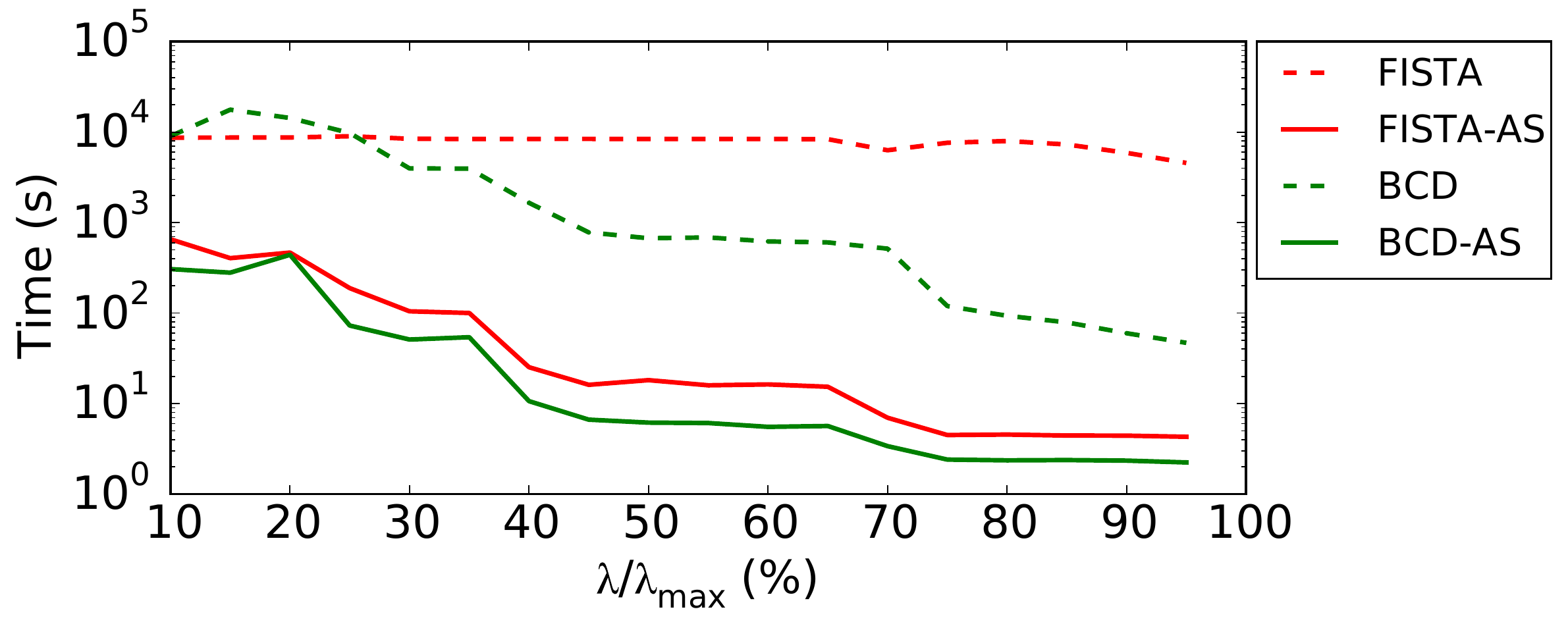}
\caption{Computation time as a function of $\lambda$ for MxNE on real MEG data (free orientation) using BCD and FISTA with (solid) and without (dashed) active set strategy.}
\label{fig:comp_time_full}
\end{figure}

We first compare the performance of the proposed BCD scheme for solving the weighted MxNE with the Fast Iterative Shrinkage Thresholding Algorithm (FISTA) \cite{Beck09Fista}, a proximal gradient method used in \cite{gramfort-etal:2012}. Both methods were applied with and without active set strategy. All computations were performed on a computer with a 2.4\,GHz Intel Core 2 Duo processor and 8\,GB RAM. The computation times as a function of $\lambda$ are presented in Fig.~\ref{fig:comp_time_full}. The BCD scheme outperforms FISTA both with and without active set strategy. Combining the BCD scheme and the active set strategy reduces the computation time by a factor of 100 and allows to compute the MxNE on real MEG/EEG data in a few seconds. Since subsequent MxNE iterations are significantly faster due to the restriction of the source space, irMxNE also runs in a few seconds on real MEG/EEG source localization problems.
\vspace{-0.2cm}

\begin{figure}[!h]
    \centering
    \includegraphics[width=0.95\columnwidth]{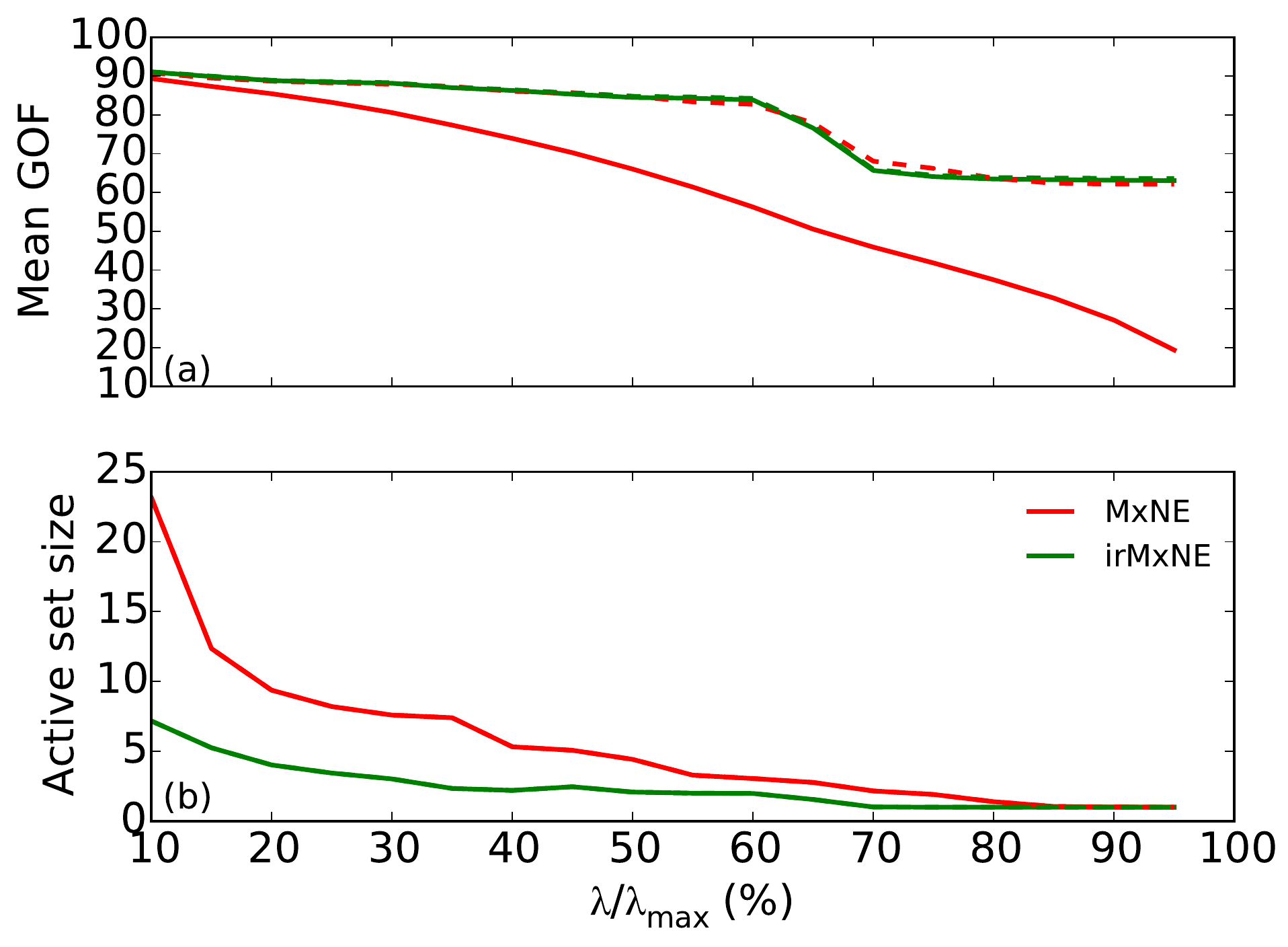}
    \caption{Mean GOF, and active set size for MxNE and irMxNE without (solid) and with (dashed) debiasing for the AEF data.}
    \label{fig:results_aud_fit}
\end{figure}

We applied MxNE and irMxNE (with and without debiasing) with different regularization parameters $\lambda$ to 100 AEF data sets generated by averaging randomly selected subsets of epochs. The mean GOF around the N100m component (from 90\,ms to 150\,ms) and the mean active set size are presented in Fig.~\ref{fig:results_aud_fit} as a function of $\lambda$.
We can see that the debiasing procedure has a strong effect on MxNE, whereas the GOF of the irMxNE result is only slightly improved indicating less amplitude bias. Debiased MxNE and irMxNE yield similar GOFs with similar plateaus, but irMxNE provides a sparser, i.e., simpler model.
\\

\begin{figure}[h]
    \centering
    \includegraphics[width=0.98\columnwidth]{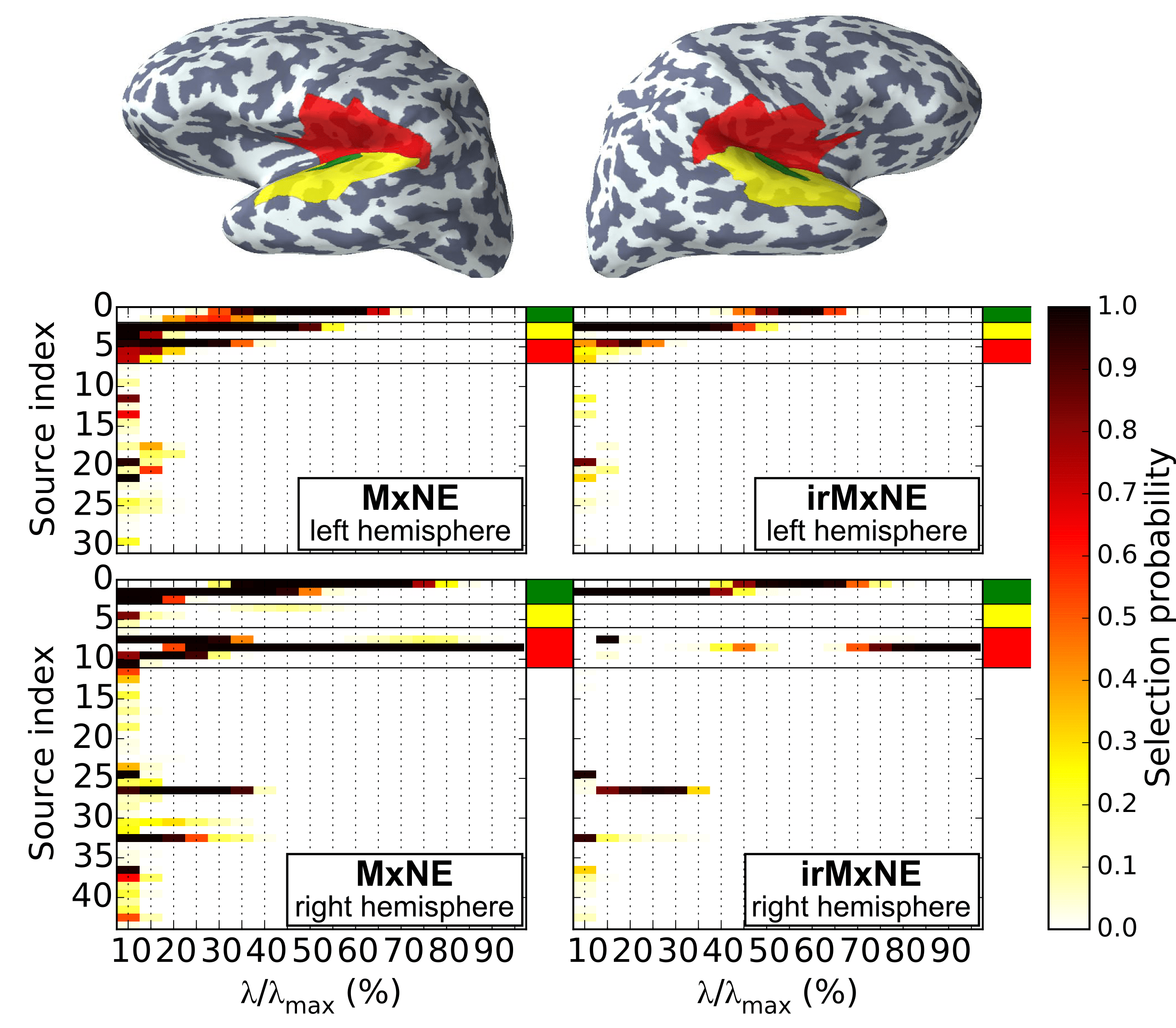}
    \caption{Source selection probability for the AEF data set using MxNE (left) and irMxNE (right). The plot is restricted to sources that are active in at least one random subsample. The colored patches on the inflated brain indicate regions of interest based on anatomical labels (green, yellow, red). Source indices, which are in the regions of interest, are highlighted by corresponding color marks. The transversal temporal gyrus is indicated in green.}
    \label{fig:results_aud_stability}
\end{figure}

The selection probability for all sources being, at least once, part of the active set obtained with MxNE or irMxNE is shown in Fig.~\ref{fig:results_aud_stability}. MxNE selects multiple sources with high probability within each region of interest, which is a consequence of the correlated design. The irMxNE approach is more selective and provides sparser source estimates. Moreover, the number of false positives, i.e., sources outside of the regions of interest, is lower for irMxNE, particularly for low values of $\lambda$.\\

\begin{figure*}[]
    \centering
    \vspace{-0.6cm}
    \begin{minipage}{0.7\linewidth}
    \begin{subfigure}[b]{0.48\linewidth}
        \includegraphics[width=\linewidth]{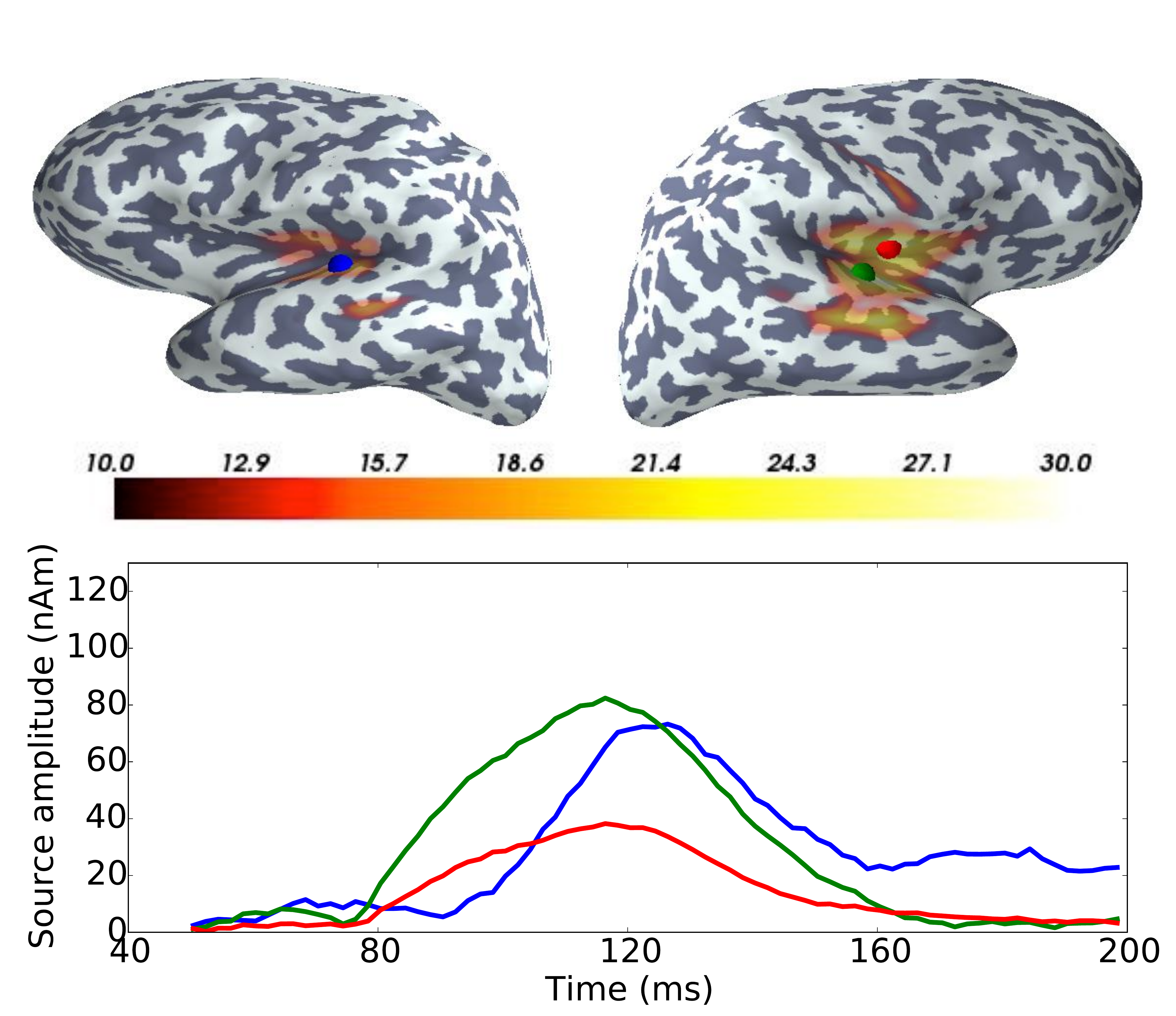}
        \caption{MxNE with $\lambda / \lambda_{max} = 60\%$}
        \label{fig:loc_aud_mxne_60}
    \end{subfigure}
    \begin{subfigure}[b]{0.48\linewidth}
        \includegraphics[width=\linewidth]{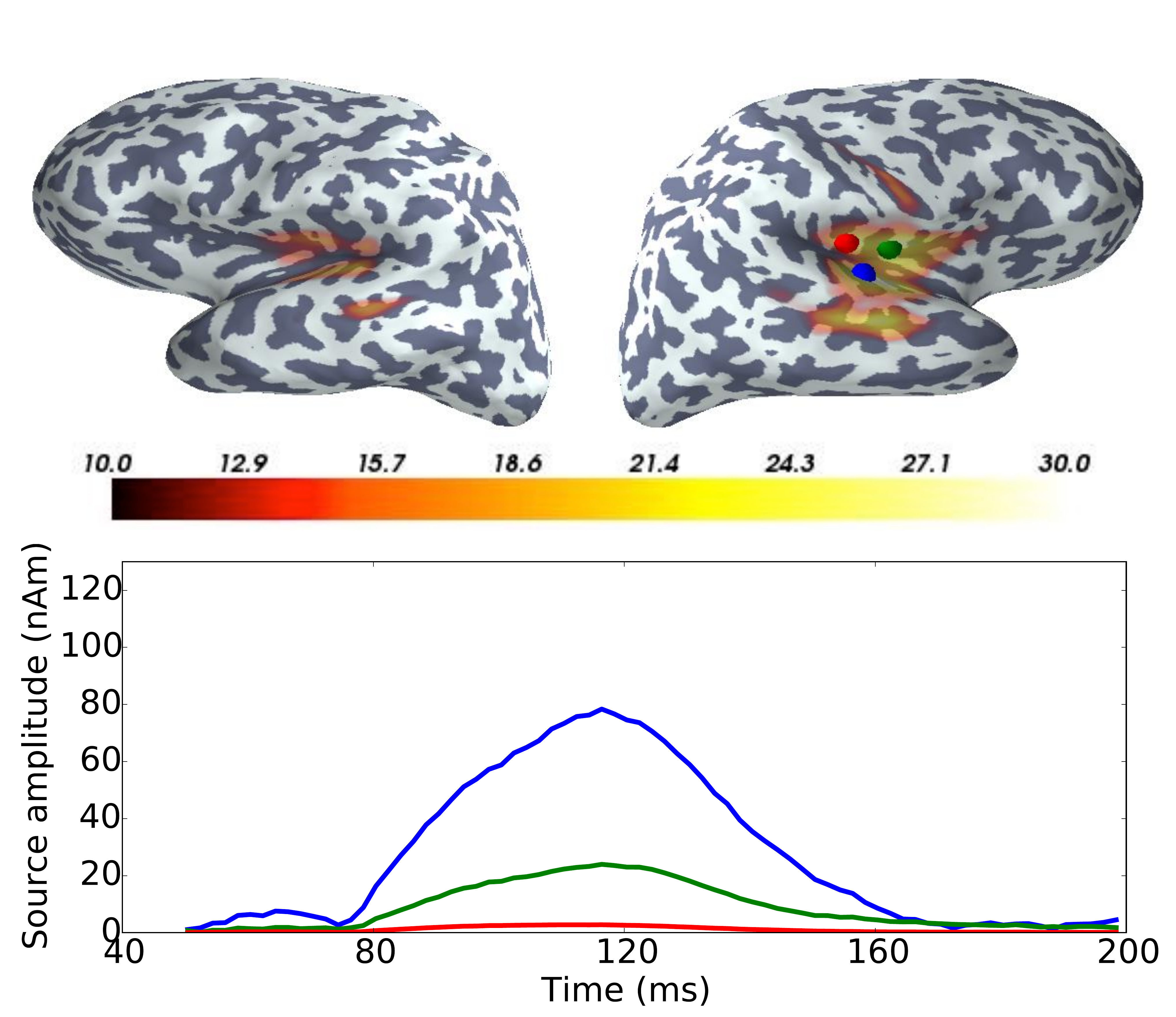}
        \caption{MxNE with $\lambda / \lambda_{max} = 70\%$}
        \label{fig:loc_aud_mxne_70}
    \end{subfigure}
    \begin{subfigure}[b]{0.48\linewidth}
        \includegraphics[width=\columnwidth]{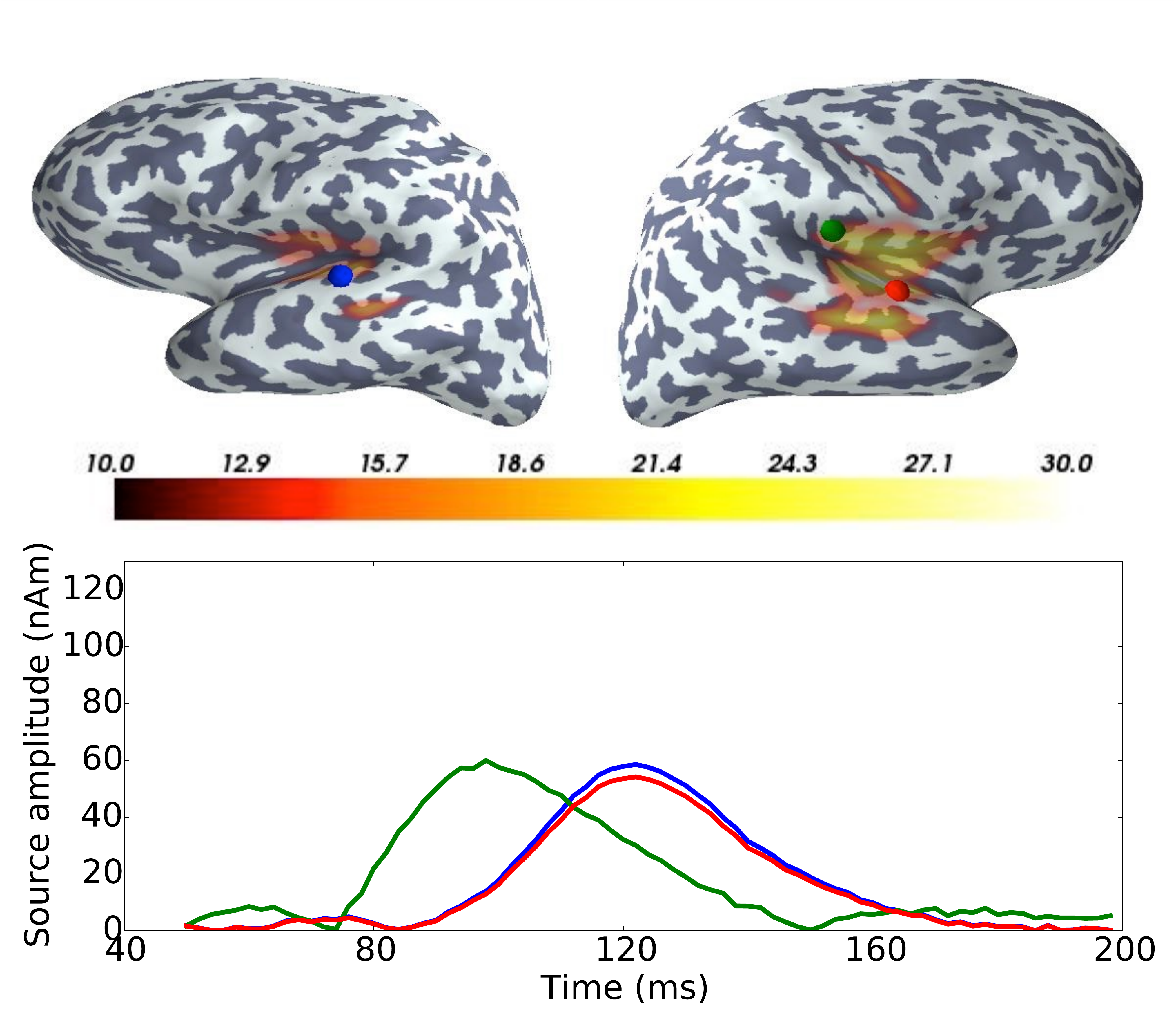}
        \caption{RAP-MUSIC (2 independent topographies)}
        \label{fig:loc_aud_rapmusic}
    \end{subfigure}
    \hspace{10pt}
    \begin{subfigure}[b]{0.48\linewidth}
        \includegraphics[width=\linewidth]{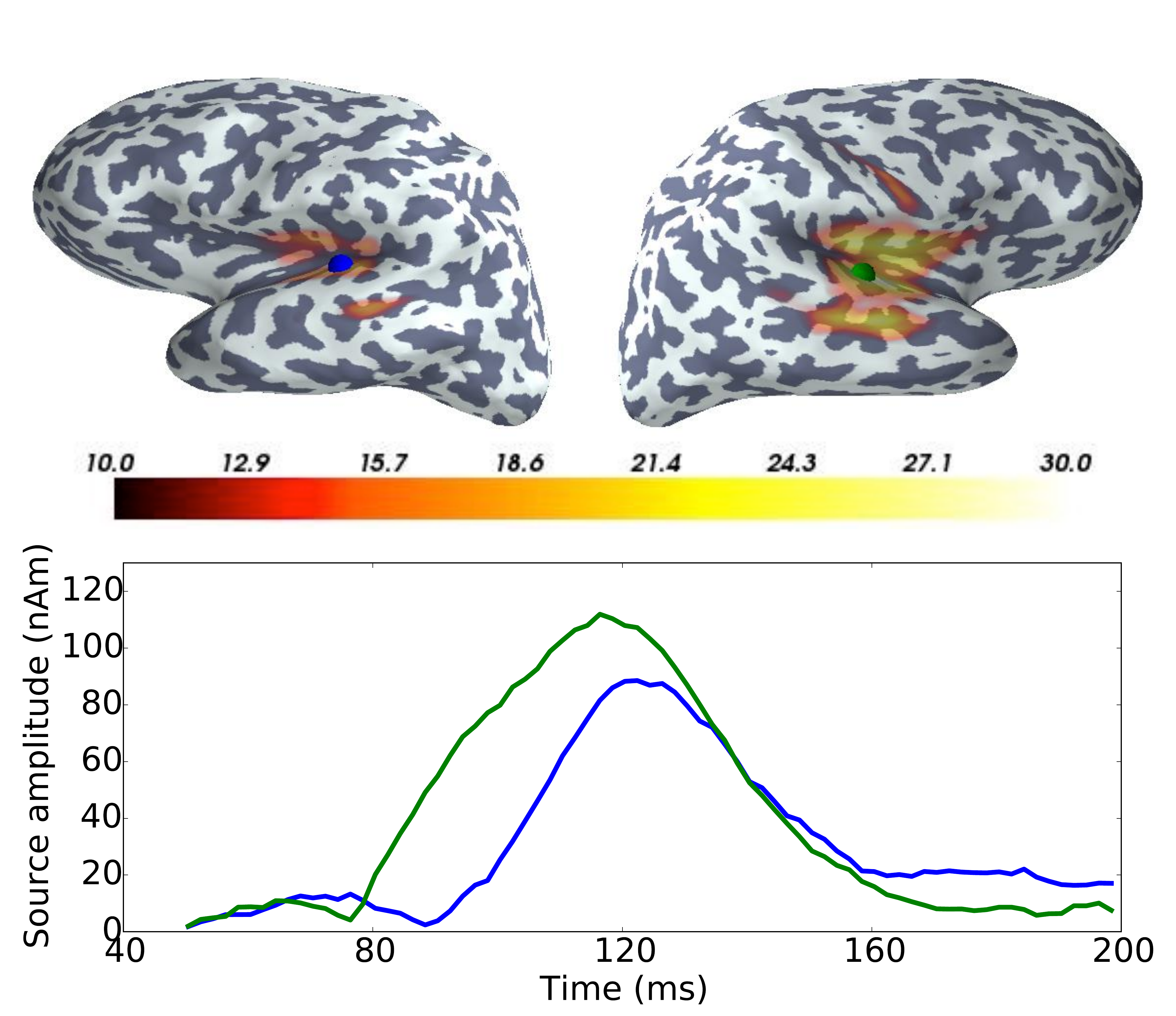}
        \caption{irMxNE with $\lambda / \lambda_{max} = 60\%$}
        \label{fig:loc_aud_irmxne_60}
    \end{subfigure}
    \end{minipage}
    \hfill
    \begin{minipage}{0.25\linewidth}
    \caption{Source reconstruction results using AEFs evoked by left auditory stimulation. The estimated source locations for MxNE (a, b), RAP-MUSIC (c) and irMxNE (d), indicated by colored spheres, and the corresponding time courses are color-coded. The maximum of the dSPM estimate per source, which is thresholded for visualization purposes, is shown as an overlay on each cortical surface.}
    \label{fig:localization_aud}
    \end{minipage}
\end{figure*}

Exemplary source reconstructions for debiased MxNE and irMxNE are illustrated in Fig.~\ref{fig:localization_aud}. For comparison, we present a RAP-MUSIC estimate based on single- and two-dipole independent topographies \cite{mosher-leahy:1999, mosher-leahy:1998}. The maximum dSPM score \cite{Dale-etal:2000} per source is shown as an overlay on each cortical surface. 
MxNE with $\lambda / \lambda_{max} = 60\%$ shows activation in both primary auditory cortices with main peaks around \mbox{110 ms} corresponding to the N100m component. The activation on the right hemisphere is however split into two highly correlated dipoles, which are partly located on the wrong side of the Sylvian fissure. Increasing $\lambda$ does not fix the latter issue, since dipoles in the left primary auditory cortex are eliminated before actually erasing spurious activity on the right hemisphere. The loss of the active source in the left auditory cortex is also indicated by the drop of the GOF in Fig.~\ref{fig:results_aud_fit}. The size of the signal subspace for RAP-MUSIC was estimated to be 50 by the thresholding procedure. Being based on an empirical estimate of the data covariance, this procedure tends to overselect the rank of the signal subspace \cite{mosher-leahy:1999} and the RAP-MUSIC estimate depends on the correlation threshold. With the setting proposed in \cite{mosher-leahy:1998}, only two independent topographies, a single- and a two-dipole topography, yield sufficient subspace correlations. The dipoles are reconstructed close to the primary auditory cortex on both hemispheres. The GOF of the three-dipole model is 86.7\%. Using single- and two-dipole topographies provides better RAP-MUSIC estimates than using only single-dipole topographies. The irMxNE with $\lambda / \lambda_{max} = 60\%$, which converged after 10 iterations, reconstructs single dipoles in both primary auditory cortices. Intuitively, the green and blue sources, which are the strongest sources according to MxNE with $\lambda / \lambda_{max} = 60\%$, are favored at the next iteration of the reweighted scheme pruning out the source on the wrong side of the Sylvian fissure present in the MxNE result. The estimated source locations roughly match the peaks of the dSPM estimate. The source estimate obtained with dSPM or similar linear inverse methods (sLORETA, MNE, etc.) is however spatially smeared. To reduce the smearing of the dSPM estimate, one could increase the threshold, yet it would make it time dependent and certainly too high to see weaker sources. Alternatively, post-processing is generally required, e.g. by defining regions of interest, to improve interpretability. Note also that, in contrast to dSPM, source amplitudes obtained with irMxNE are moments of electrical dipoles expressed in nAm, which is similar to dipole fitting procedures~\cite{scherg-etal:85}. The GOF of the two-dipole model obtained with irMxNE is 81.9\% and thus only slightly lower than the three-dipole model obtained with RAP-MUSIC.\\

\subsubsection{Somatosensory evoked fields}
\begin{figure}[h]
    \centering
    \includegraphics[width=0.95\columnwidth]{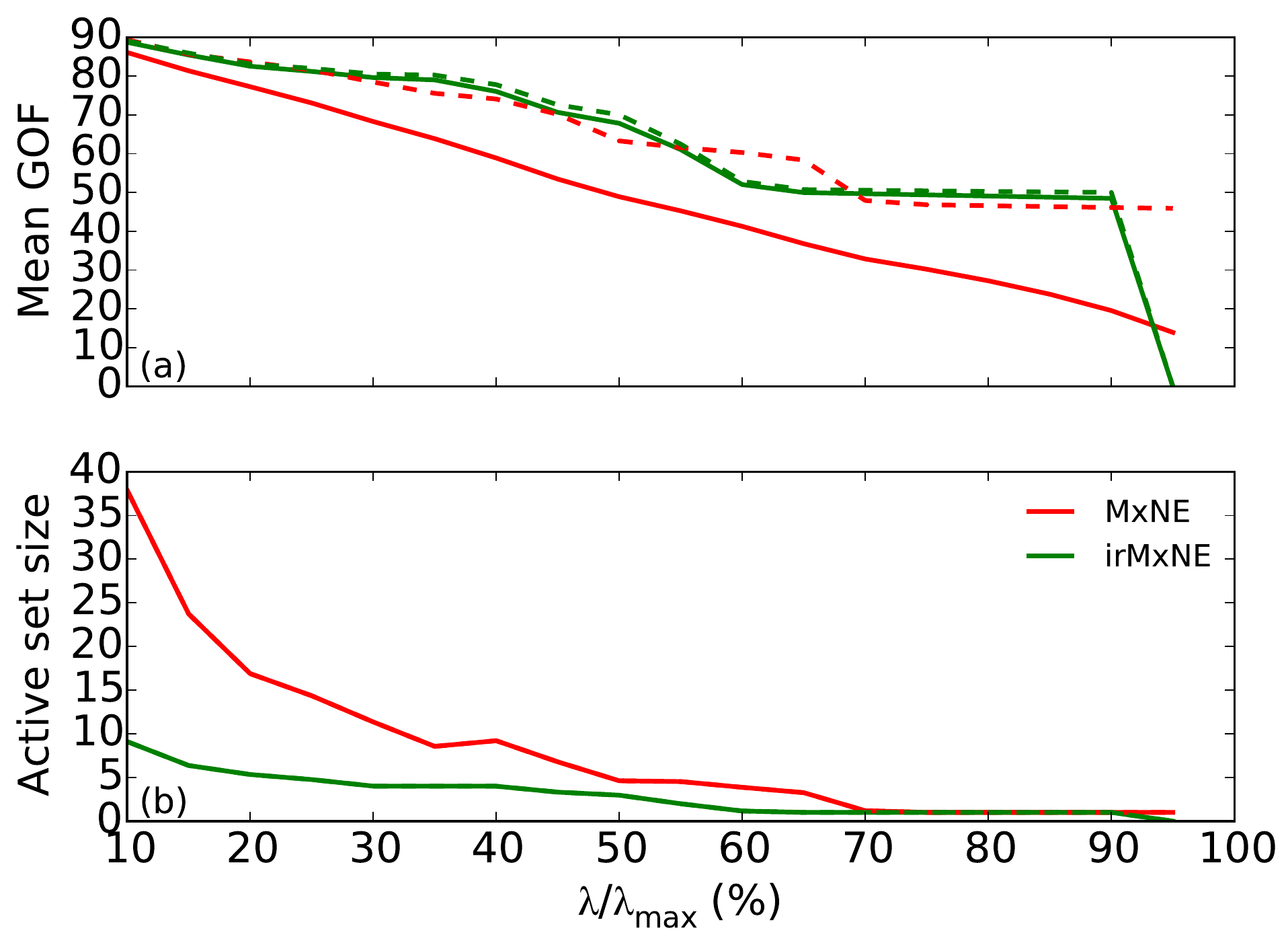}
    \caption{Mean GOF, and active set size for MxNE and irMxNE without (solid) and with (dashed) debiasing for the SEF data.}
    \label{fig:results_sef_fit}
\end{figure}

We applied MxNE and irMxNE (with and without debiasing) with different regularization parameters to 100 averaged random subsets of epochs of the SEF data set. The mean GOF and the corresponding active set size for MxNE and irMxNE (with and without debiasing) as a function of the regularization parameter $\lambda$ are shown in Fig.~\ref{fig:results_sef_fit}. We can see again that irMxNE yields significantly sparser source estimates, which however allow for a better GOF compared to MxNE. The GOFs of the MxNE and irMxNE results with and without debiasing illustrate also that the irMxNE source estimates are less biased in amplitude.
\\

\begin{figure}[!h]
    \centering
    \includegraphics[width=0.98\columnwidth]{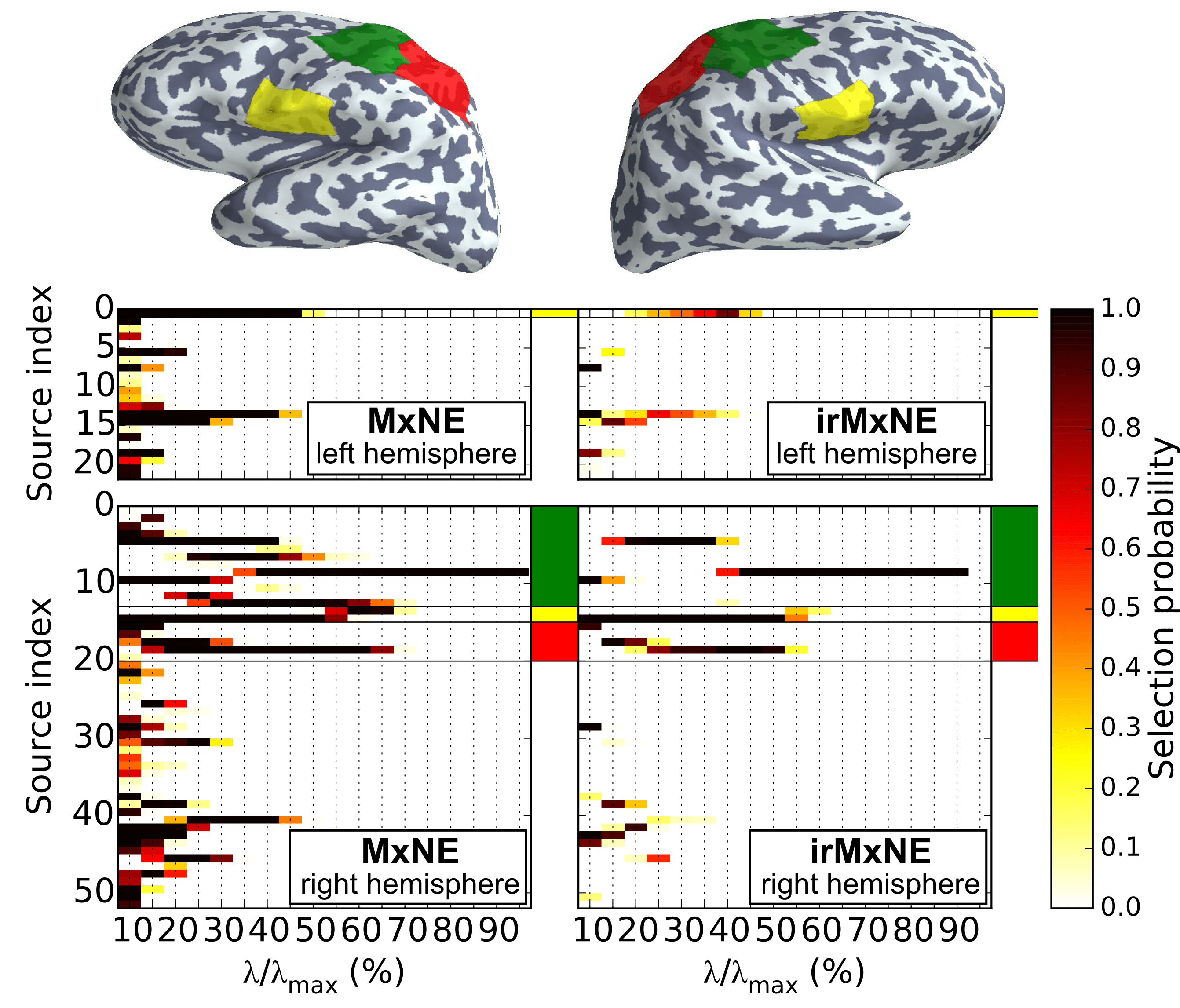}
    \caption{Selection probability for sources obtained with MxNE (left) and irMxNE (right) for the SEF data. The plot is restricted to sources that are active in at least one random subsample. The colored patches on the inflated brain indicate regions of interest based on anatomical labels (green, yellow, red). Source indices, which are in the regions of interest, are highlighted by corresponding color marks.}
    \label{fig:results_sef_stability}
\end{figure}

Fig.~\ref{fig:results_sef_stability} presents the selection probability for all sources, which are non-zero in at least one MxNE or irMxNE estimate. The irMxNE typically selects only one source per region of interest for different values of $\lambda$. The number of false positives is also significantly lower. These results confirm the findings obtained from the AEF data set in section \ref{sec:realdata_aud}. The stability analysis reveals also that the source in the ipsilateral secondary somatosensory cortex (iS2) is less stable compared to the contralateral sources, which might be caused by its relatively weak field pattern \cite{Sorrentino-etal:2009}.

\begin{figure*}[]
    \centering
    \vspace{-0.6cm}
    \begin{minipage}{0.7\linewidth}
    \begin{subfigure}[b]{0.48\linewidth}
        \includegraphics[width=\linewidth]{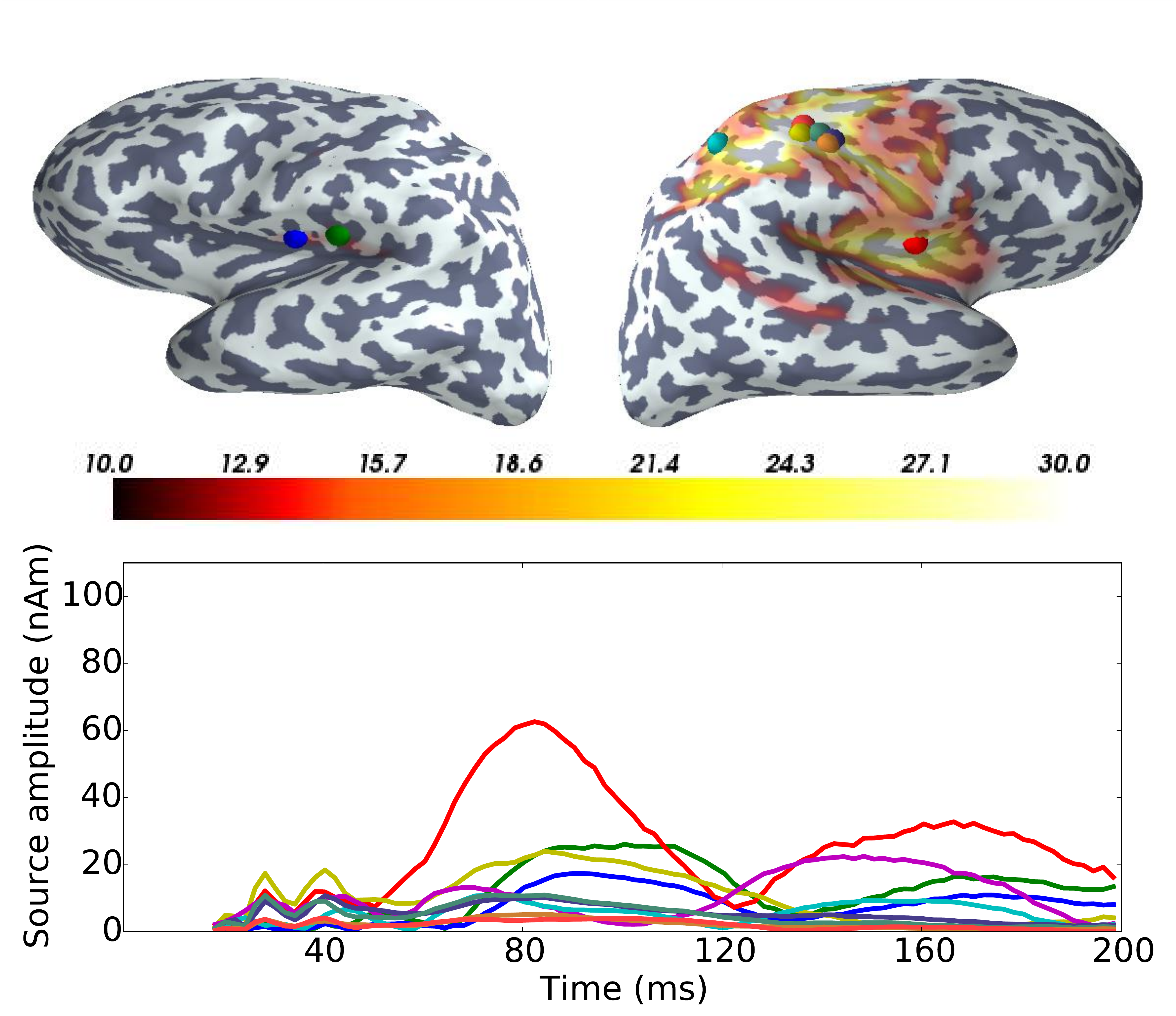}
        \caption{MxNE with $\lambda / \lambda_{max} = 40\%$}
        \label{fig:loc_sef_mxne_40}
    \end{subfigure}
    \begin{subfigure}[b]{0.48\linewidth}
        \includegraphics[width=\linewidth]{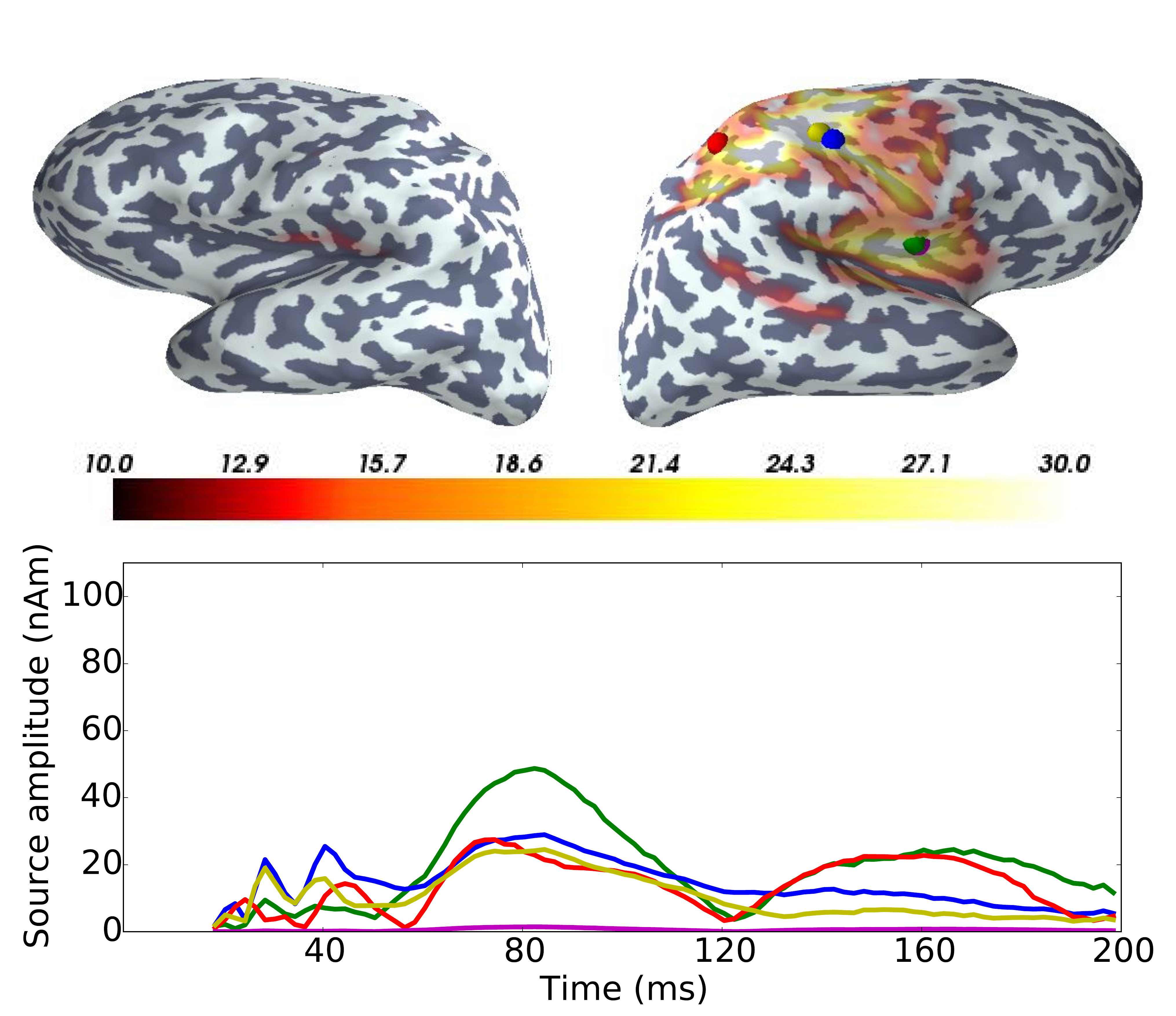}
        \caption{MxNE with $\lambda / \lambda_{max} = 55\%$}
        \label{fig:loc_sef_mxne_50}
    \end{subfigure}
    \begin{subfigure}[b]{0.48\linewidth}
        \includegraphics[width=\linewidth]{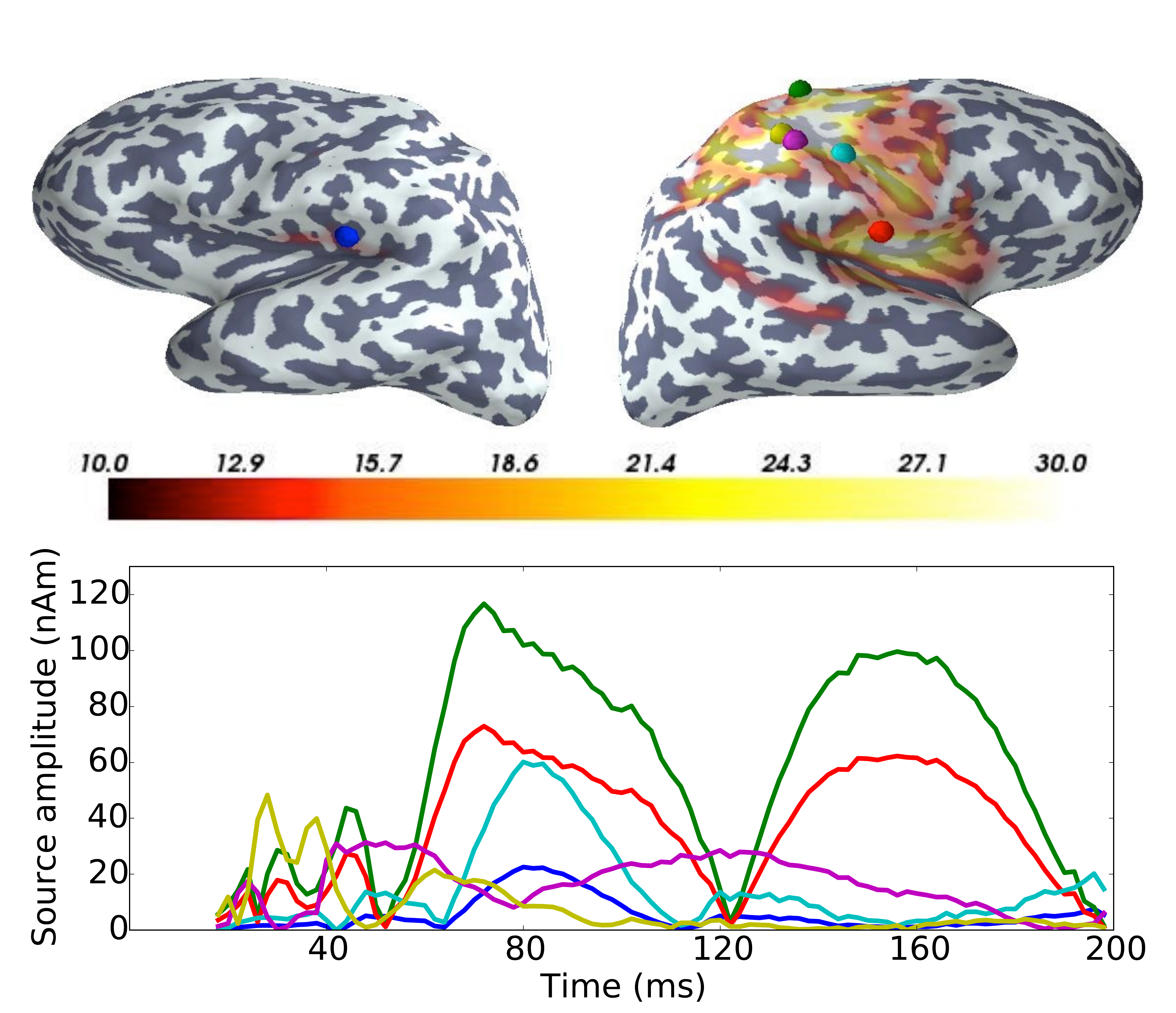}
        \caption{RAP-MUSIC (4 independent topographies)}
        \label{fig:loc_sef_rapmusic}
    \end{subfigure}
    \hspace{10pt}
    \begin{subfigure}[b]{0.48\linewidth}
        \includegraphics[width=\linewidth]{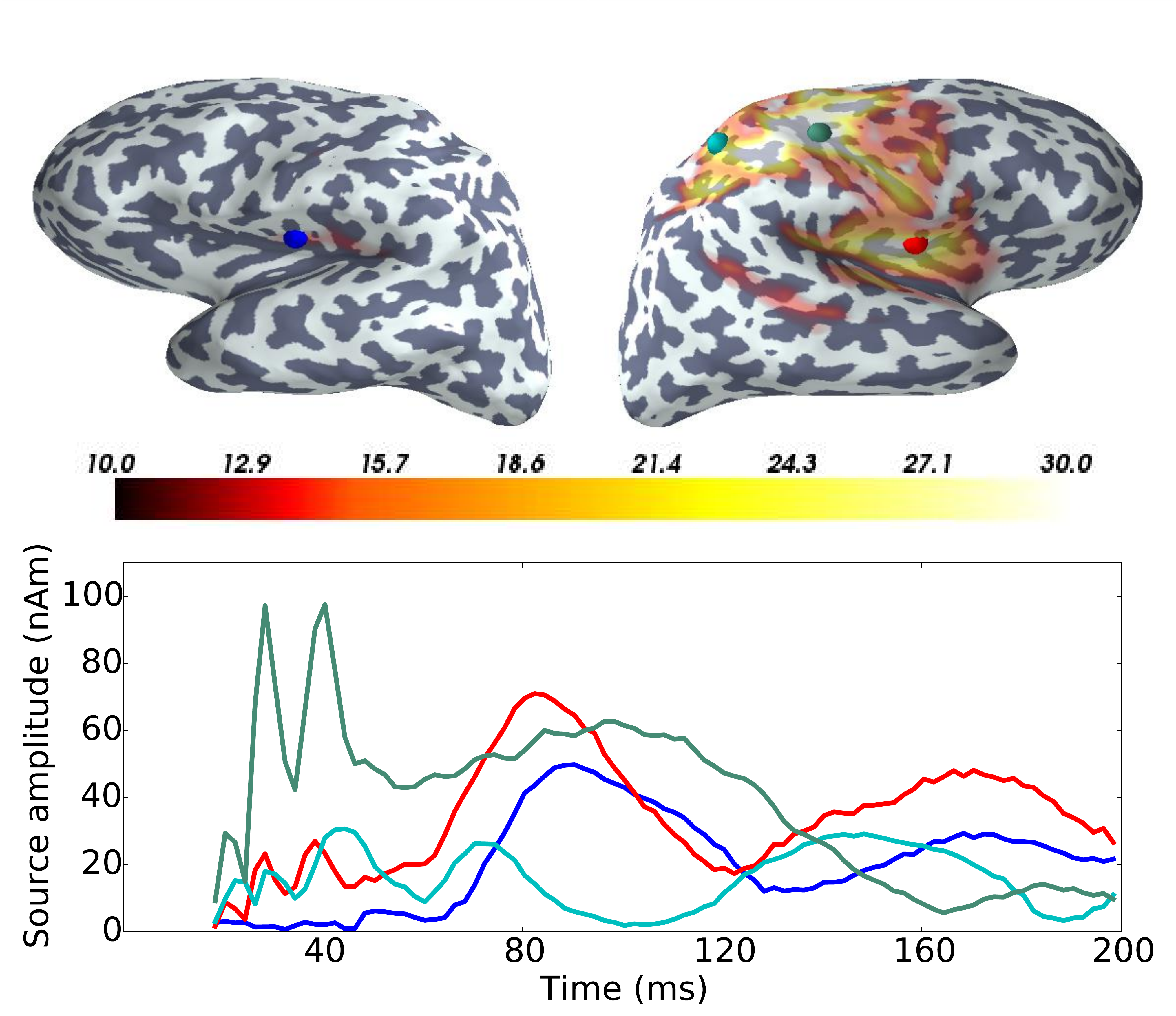}
        \caption{irMxNE with $\lambda / \lambda_{max} = 40\%$}
        \label{fig:loc_sef_irmxne}
    \end{subfigure}
    \end{minipage}
    \hfill
    \begin{minipage}{0.25\linewidth}
    \caption{Source reconstruction results using SEFs evoked by electrical stimulation of the left median nerve. The estimated source locations for MxNE (a, b), RAP-MUSIC (c) and \mbox{irMxNE (d)}, indicated by colored spheres, and the corresponding time courses are color-coded. The maximum of the dSPM estimate per source, which is thresholded for visualization purposes, is shown as an overlay on each cortical surface.}
    \label{fig:localization_sef}
    \end{minipage}
\end{figure*}

Fig.~\ref{fig:localization_sef} presents source reconstruction results obtained with MxNE and irMxNE for selected regularization parameters. 
As in section~\ref{sec:realdata_aud}, we show a RAP-MUSIC estimate and added the maximum dSPM score per source as an overlay to all subfigures. MxNE with \mbox{$\lambda / \lambda_{max} = 40\%$} reconstructs dipoles in the contralateral primary somatosensory cortex (cS1), the contralateral and ipsilateral secondary somatosensory cortices (cS2 and iS2), and the posterial parietal cortex (cPPC). The source locations roughly coincide with the main peaks of the dSPM estimate. As for the AEF data set, the source activation per region is   split into several correlated dipoles. An increase of the regularization parameter results in a loss of physiologically meaningful source activity such as activation in iS2, which is visible in the dSPM estimate. The relevance of this activation is also indicated by the drop of the GOF in Fig.~\ref{fig:results_sef_fit}. The signal subspace estimation for RAP-MUSIC yields a signal subspace size of 43. Hence, the RAP-MUSIC estimate depends on the choice of the correlation threshold. For our settings, four independent topographies, two single- and two two-dipole topographies, are above the subspace correlation threshold. Dipoles are reconstructed in all relevant areas. The activation in cS1 is split into several dipoles, probably due to the fixed orientation source model applied in RAP-MUSIC. The GOF of this six-dipole model is 82,6\%. The RAP-MUSIC estimate benefits from using single and two-dipole topographies. The irMxNE approach with \mbox{$\lambda / \lambda_{max} = 40\%$} converged after 14 reweightings. The resulting source estimate contains four single dipoles representing activation in each of the four regions. The GOF of the four-dipole model obtained with irMxNE is 81.4\% and thus higher than the GOF of the corresponding MxNE estimate and only slightly lower than the GOF of the six-dipole model obtained with RAP-MUSIC.

\section{Discussion and Conclusion}
In this work, we presented irMxNE, an MEG/EEG inverse solver based on regularized regression with a non-convex block-separable penalty. The non-convex optimization problem is solved by iteratively solving a sequence of weighted MxNE problems, which allows for fast algorithms and global convergence control at each iteration. We proposed a new algorithm for solving the MxNE surrogate problems combining BCD and a forward active set strategy, which significantly decreases the computation time compared to the original MxNE algorithm \cite{gramfort-etal:2012}. This new algorithm makes the proposed iterative reweighted optimization scheme applicable for practical MEG/EEG applications. The approach is also applicable to other block-separable non-convex penalties such as the logarithmic penalty proposed in \cite{Candes} by adapting the definition of the weights in Eq.~\eqref{eqn:comp_weights}. The irMxNE method is designed for offline source reconstruction, which is still the main application of MEG/EEG source imaging in research and clinical routine. However, we are aware of a growing interest in real-time brain monitoring \cite{dinh:2015}. New techniques such as parallel BCD schemes \cite{li-osher:2009}, clustering approaches \cite{dinh-etal:2015}, and safe rules \cite{fercoq-etal:2015} can help to further reduce the computation time. As proposed in \cite{Candes, Gasso}, the first iteration of irMxNE is equivalent to computing the standard MxNE. Consequently, the irMxNE result is at least as sparse as the MxNE estimate. The iterative reweighting procedure can thus be considered as a post-processing for MxNE improving source recovery, stability, and amplitude bias. This was confirmed by empirical results based on simulations and two MEG data sets. We attribute this to the spatial correlation and the poor conditioning of the forward operator in MEG/EEG source analysis. An alternative approach to improve the conditioning of the inverse problem based on clustering the columns of the gain matrix is presented in \cite{dinh-etal:2015}, which however affects the spatial resolution. The source locations reconstructed by irMxNE roughly coincided with the main peaks of the dSPM estimate, which demonstrate that the proposed inverse solver can present a simple and easy-to-interpret spatio-temporal picture of the active sources. The models reconstructed with RAP-MUSIC provided a slightly higher goodness of fit, but contained more active sources. We found that RAP-MUSIC benefits from using single-dipole and two-dipole independent topographies. The use of higher-order source models, which are limited to a small number of correlated sources, however significantly increases the computational complexity, particularly for source spaces with high resolution. Approaches improving the computation time of RAP-MUSIC are presented in \cite{dinh:2015}. In contrast, irMxNE makes no assumption on the number of correlated sources. Its computation time is not dramatically affected by the resolution of the source space. Model selection for sparse source imaging approaches, which amounts here to choosing the regularization parameter, is a critical aspect. Automatic approaches based on minimizing the prediction error such as cross-validation tend to overestimate the number of active sources and increase the false positive rate. Here, we selected the regularization parameter based on the GOF and the size of the active set. A similar procedure is used e.g. in sequential dipole fitting. The development of an automatic model selection procedure for the proposed inverse solver is future work. In particular, approaches maximizing model stability are an interesting alternative \cite{lim-yu:2015, sun-etal:2013}. Model selection is however a general issue in MEG/EEG source reconstruction. In our comparison with RAP-MUSIC, we found e.g. that the size of the signal subspace and the correlation threshold have a strong influence on the final source estimate. Due to the limited number of samples, we can only obtain an empirical estimate of the data covariance, which involves the risk of overestimating the size of the signal subspace using the thresholding procedure. The correlation threshold is used to switch to higher order source models and to account for noise components in the signal subspace. Similar to MxNE, irMxNE assumes that the locations of active sources is constant over time. Hence, it should be applied to data, for which this model assumption is approximately true, e.g., by selecting intervals of interest or applying a moving window approach. To go beyond stationary sources, the reconstruction of non-stationary focal source activation can be improved by applying sparsity constraints in the time-frequency domain such as in the TF-MxNE \cite{gramfort-etal:2013}. Preliminary results on the application of non-convex regularization for such models based on iterative reweighting procedures were presented in \cite{strohmeier-etal:2015}. The irMxNE solver is available in the MNE-Python package~\cite{mne-python}.

\section*{Acknowledgment}
The authors would like to thank M. S. H\"am\"al\"ainen for providing the experimental MEG data sets.

\bibliographystyle{IEEEtran}
\balance
\bibliography{biblio}

\end{document}